\documentclass[12pt]{article}
\usepackage[letterpaper, left=.8in, top=0.9in, right=.8in, bottom=0.70in,nohead,includefoot, verbose, ignoremp]{geometry}
\usepackage{charter} 
\usepackage{enumerate} 

\usepackage{latexsym,amssymb,amsmath,amsfonts,graphicx,color,amsthm,enumerate} 
\usepackage[svgnames,dvipsnames,x11names]{xcolor}

\usepackage[round]{natbib}
\usepackage{bibentry}
\nobibliography*

\usepackage{bm}
\usepackage{booktabs}
\usepackage{multirow}

\usepackage{hyperref} 
\usepackage{algorithmic} 
\usepackage{algorithm}

\begin{document}
\title{Sequential Bayesian Predictive Synthesis}
\author{
Riku Masuda\footnote{Graduate School of Economics, The University of Tokyo. 
\newline{
E-Mail: rikumasuda520@gmail.com}} \
and 
Kaoru Irie\footnote{Corresponding author. Faculty of Economics, The University of Tokyo. 
\newline{
E-Mail: irie@e.u-tokyo.ac.jp}}
}
\maketitle
\begin{abstract}
Dynamic Bayesian predictive synthesis is a formal approach to coherently synthesizing multiple predictive distributions into a single distribution. In sequential analysis, the computation of the synthesized predictive distribution has heavily relied on the repeated use of the Markov chain Monte Carlo method. The sequential Monte Carlo method in this problem has also been studied but is limited to a subclass of linear synthesis with weight constraint but no intercept. In this study, we provide a custom, Rao-Blackwellized particle filter for the linear and Gaussian synthesis, supplemented by timely interventions by the MCMC method to avoid the problem of particle degeneracy. In an example of predicting US inflation rate, where a sudden burst is observed in 2020-2022, we confirm the slow adaptation of the predictive distribution. To overcome this problem, we propose the estimation/averaging of parameters called discount factors based on the power-discounted likelihoods, which becomes feasible due to the fast computation by the proposed method. 

\par\vspace{4mm}
\end{abstract}

\section{Introduction}

In modern forecasting problems, it is common that multiple predictive distributions are available for an statistician. 
The source of these predictive distributions can include human experts and deterministic computer algorithms, but in data analysis, it is most likely statistical models analyzed by the statistician. 
Predictive synthesis is a problem where the statistician uses those predictive distributions to create a new, single predictive distribution for an improved prediction. 
A formal approach to this problem is Bayesian predictive synthesis (BPS); it is based on the series of research on the coherent rule of updating the statistician's predictive distribution as ``posterior'' informed by the predictive distributions as ``data'' (\citealt{genest1985modeling,westcrosse1992,West1992d}) and includes the traditional Bayesian model averaging as a special case. Its extension to dynamic models has been proposed by \cite{mcalinn2019dynamic} (dynamic Bayesian predictive synthesis, DBPS). Predictive analysis by (D)BPS has seen various applications, including the prediction of macroeconomic indices \citep{mcalinn2020multivariate,mcalinn2021mixed}, oil prices \citep{aastveit2023quantifying}, real estate prices \citep{cabel2022spatially} and decision analysis \citep{TallmanWest2022}. 

Although the synthesized prediction by DBPS outperforms other predictive methods in many scenarios (e.g., see \citealt{mcalinn2019dynamic}), its computation relies heavily on the repeated use of the Markov chain Monte Carlo (MCMC) method. In fact, in all the papers on DBPS cited above, the synthesized predictive distributions are computed by re-doing the MCMC method. This is because the synthesized predictive distribution is not analytically tractable, even if we assume the linear and Gaussian synthesis structure. 
The computation cost becomes more expensive if the statistician has to compute the synthesis of the predictive distributions multiple times. Examples include real-time monitoring, sequential forecasting, and calibrating the predictive synthesis (or trying different ways of synthesizing the predictive distributions). 
Especially in sequentially processing the streaming data, such repeated use of the MCMC method is prohibitive, which has hindered the use of DBPS in practice. In this context, seeking online computation by the sequential Monte Carlo (SMC) method is natural. 

In this study, we propose an SMC-based approach customized for DBPS and evaluate its computational performance. 
The use of the particle filter \citep{gordon1993novel} has been considered in the context of forecast density pooling/combination \citep{billio2013time,casarin2015parallel}, and has been limited to a particular class of synthesis functions. 
We, too, consider the particle filter, but extend its application to the general class of synthesis functions in the DBPS framework. 
Furthermore, we focus on the linear and Gaussian synthesis and provide the custom, Rao-Blackwellized Bootstrap particle filter for online predictive analysis (\citealt{chen2000mixture,doucet2000sequential}). 
To handle the particle degeneracy problem, we monitor the effective sample size (ESS) and, if the ESS is small and below a threshold, we intervene in the sequential computation using the off-line MCMC method instead. 
An appropriate and timely intervention by the MCMC method is crucial for the accurate computation of the synthesized predictive distribution and does not critically slow down the sequential computation. 

The proposed method--- the pair of the DBPS model and filtering algorithm--- is successful in providing almost the same predictive distribution as the MCMC method does in most cases. 
However, the DBPS is critically insensitive to a sudden burst of observations (or potential structural change) and could perform poorly in predictions, regardless of the computational methods. 
This sudden burst is typical in the analysis of inflation rates studied in the literature, especially in the worldwide pandemic and economic crisis in 2020-2022. 
To tackle this problem, we introduce the loss discounting framework (LDF, \citealt{bernaciak2022loss}) into the DBPS. This framework requires the computation of the DBPS multiple times at every time point, which is a typical example where the fast posterior/predictive computation by the proposed SMC method is crucial. 

The rest of this paper is organized as follows. Section~\ref{sec:DBPS} reviews the DBPS and the linear and Gaussian synthesis functions, emphasizing properties related to the SMC method. In Section~\ref{sec:smc}, we develop the particle filter for the DBPS, including its Rao-Blackwellization. We illustrate the proposed SMC method using the dataset of the US inflation rates in Section~\ref{sec:data}, where the efficiency and predictive performance of the proposed method are evaluated. In Section~\ref{sec:discount}, we discuss the introduction of the LDF into the DBPS and observe the improved predictive performance in the inflation example. We comment on future research in Section~\ref{sec:future}

\ 

\textbf{Notational remarks:} We denote the normal distribution with mean vector $\mu$ and variance matrix $\Sigma$ by $N(\mu, \Sigma)$. For integers $s<t$, we write $s:t = (s,s+1,\dots, t-1,t)$.

\section{Dynamic Bayesian predictive synthesis} \label{sec:DBPS}

\subsection{Overview and terminology}

Suppose that $K$ predictive distributions, $\mathcal{H}_t = \{ h_{1t},\dots, h_{Kt}  \}$, are provided to forecast quantity $y_t$. Bayesian predictive synthesis (or density combination if density functions exist) discusses how we should combine the $K$ predictive distributions into a single distribution as the final output for prediction. The source of the predictive distributions can be anything, including reports from human experts (e.g., \citealt{chernis2022combining}). However, in most applications of the BPS, those predictive distributions are obtained from statistical models. We call those $K$ models and predictive distributions {\it agent models} and {\it agent predictive distributions}, respectively. The resulting predictive distribution of the BPS is named {\it synthesized} predictive distribution. In this study, we focus on the one-step ahead prediction for simplicity. Still, the BPS is open to multi-step ahead predictions and path forecasting, to which our study can be easily generalized.  

The dynamic BPS involves two steps of computation: model learning and synthesis calibration. The former is the process of creating agent predictions $\mathcal{H}_t$, and the latter is to revise (or {\it calibrate}) the way of synthesizing $\mathcal{H}_t$. Since the calibration step depends on $\mathcal{H}_t$ created by learning about the agent models, these steps cannot be processed in parallel.

\subsection{Agent predictive models}

Let $\{ y_t \} _{t=1:T}$ be univariate time series to be forecast. We have $K$ agent models about $y_t$, denoted by $\{ \mathcal{M}_k \} _{k=1:K}$. At each time point $t-1$, we learn about each model $\mathcal{M}_k$ by using observed $y_{1:(t-1)}$ and compute the one-step ahead agent predictive distribution, namely, $p(  y_t | y_{1:(t-1)}, \mathcal{M}_k )$. We denote this agent predictive density by $h_{kt}(\cdot)$ as 
\begin{equation*}
    h_{kt}(x_{kt}) = p( x_{kt} | y_{1:(t-1)}, \mathcal{M}_k),
\end{equation*}
where we use $x_{kt}$ as the argument of the density, not $y_t$, to emphasize that this is the prediction made by agent model $\mathcal{M}_k$ at time $t-1$. The set of agent predictive densities, $\mathcal{H}_t$, becomes available at time $t-1$.

In this study, we use the conjugate dynamic models as the agent models whose online posterior and predictive distributions are obtained analytically. In general, the agent predictive distributions are not analytically tractable, and are often approximated using simulation-based methods (SMC, MCMC, and others). The computational methods we discuss below are also applicable to such approximated predictive distributions, where the simulation from (appoximated) $h_{kt}$ is feasible.

\subsection{BPS, synthesis functions and priors}
A formal approach to synthesizing the $K$ agent predictive distributions is to use the conditional distribution $p(y_t | \mathcal{H}_t )$, which is difficult to compute directly using the Bayes rule. 
Instead, in dynamic Bayesian predictive synthesis (DBPS, \citealt{mcalinn2019dynamic}), we synthesize the $K$ agent predictive densities, $\mathcal{H}_t = \{ h_{1t},\dots, h_{Kt}  \}$, by time-varying synthesis function $\alpha_t (y_t | x_t)$ at each time as follows: 
\begin{equation*}
    p(y_t | \mathcal{H}_{t} , \alpha_t) = \int \alpha_t(y_t | x_t) h_t (x_t) dx_t,
\end{equation*}
where $x_t = (x_{1t} , \cdots , x_{Kt})^\prime$ and $h_t(x_t) = \prod_{k=1}^K h_{kt}(x_{kt})$. In fact, the distribution of $y_t$ conditional on $\mathcal{H}_t$ {\it must} be of this form for the joint, marginal, and conditional distributions to be consistently defined  (\citealt{genest1985modeling}, \citealt{westcrosse1992}, \citealt{West1992d}). 
In practice, the synthesis function, $\alpha_t(\cdot )$, must be specified by users. We parametrize this synthesis function by using dynamic {\it calibration} parameters, $\{\Phi_t\}_{t = 1:T}$, and rewrite the synthesis equation as 
\begin{equation*}
    p(y_t | \mathcal{H}_{t} , \Phi_t) = \int \alpha(y_t | x_t , \Phi_t) h_t (x_t) dx_t ,
\end{equation*}
where the functional form of $\alpha (\cdot )$ is pre-determined and unchanged over time. 
Note that, in predicting $y_t$, we condition not only set of agent predictions $\mathcal{H}_{1:t}$ but also past observations $y_{1:(t-1)}$. 
Thus, the synthesized predictive distribution, or the final output for prediction, is obtained by marginalizing $\Phi_t$ out as 
\begin{equation*}
\begin{split}
    p(y_t | \mathcal{H}_{1:t} , y_{1:(t-1)}) &= \int p(y_t | \mathcal{H}_{t} , \Phi_t)p(\Phi_t | y_{1:(t-1)} , \mathcal{H}_{1:(t-1)}) d\Phi_t \\
    &= \iint \alpha(y_t | x_t , \Phi_t) h_t (x_t)p(\Phi_t | \mathcal{H}_{1:(t-1)} ,  y_{1:(t-1)}) dx_t d\Phi_t .
\end{split}
\end{equation*}
As seen in this expression, we calibrate how to synthesize the multiple agent predictive densities by learning about the calibration parameter, $\Phi_t$. 
The set of calibration parameters includes the location and scale of each forecast and adjusts the bias of the point prediction and predictive uncertainty, as specified in Section~\ref{subsec:DBPS} in detail. 

To compute $p(y_t | \mathcal{H}_{1:t} , y_{1:(t-1)})$ using the integral expression above, we need to calculate $p(\Phi_t | \mathcal{H}_{1:(t-1)} ,  y_{1:(t-1)})$, or the online prior distribution of calibration parameters. 
This distribution is further augmented as 
\begin{equation*}
\begin{split}
    &p(\Phi_t  | \mathcal{H}_{1:(t-1)} , y_{1:(t-1)}) \\
    &= \int p(\Phi_t | \Phi_{1:(t-1)}, x_{1:(t-1)} , y_{1:(t-1)} ) p(\Phi_{1:(t-1)}, x_{1:(t-1)} | \mathcal{H}_{1:(t-1)} ,  y_{1:(t-1)} ) d\Phi_{1:(t-1)}dx_{1:(t-1)}.
\end{split}
\end{equation*}
Here, the evolution of the calibration parameters, $p(\Phi_t | \Phi_{1:(t-1)}, x_{1:(t-1)} , y_{1:(t-1)} )$, depends on the specification of synthesis function $\alpha$ and discussed in the next subsection. The SMC method is applied to the computation of the online posterior distribution of $(\Phi _{1:(t-1)},x_{1:(t-1)})$ and studied in Section~\ref{sec:smc}.

\subsection{Dynamic linear model synthesis} \label{subsec:DBPS}
Dynamic linear models (DLMs), or linear and Gaussian state space models, are an important class of dynamic models (e.g., see \citealt{prado2021time}) and can also be used in modeling synthesis function $\alpha (y_t|x_t,\Phi_t)$ and the evolution of $\Phi_t$. Following \cite{mcalinn2019dynamic}, we define the DLM synthesis with discount factors $\beta , \delta \in (0,1]$ as 
\begin{equation*}
    \begin{split}
        & \alpha(y_t | x_t , \Phi_t) = N(y_t | F_t^\prime \theta_t , \nu_t )\\
        & \theta_t = \theta_{t-1} + \omega_t \ , \ \omega_t \sim N(0 , \nu_t W_t)\\
        & \nu_t = \frac{\beta}{\gamma_t}\nu_{t-1} \ , \ \gamma_t \sim Be\left( \beta \frac{n_{t-1}}{2} , (1-\beta)\frac{n_{t-1}}{2} \right)
    \end{split}
\end{equation*}
where $F_t = (1 , x_{1t} , \cdots , x_{Kt})^\prime$, $\theta_t = (\theta_{0t} , \theta_{1t} \cdots , \theta_{Kt})^\prime$, and $Be(a,b)$ denotes beta distribution. State variance $W_t$ is defined via a standard, single discount factor specification with state evolution discount factor $\delta \in (0,1]$. The residual variance $\nu_t$ follows the standard beta-gamma random walk volatility model with discount factor $\beta \in (0,1
]$ where the shape parameter is updated by $n_{t} = \beta n_{t-1} + 1$. In this setting, online posterior $p(\theta_t,\nu_t | x_{1:t} , y_{1:t})$ and prior $p(\theta_{t+1},\nu_{t+1} | x_{1:t} , y_{1:t})$ are normal-inverse-gamma distributions whose sufficient statistics are deterministically updated. Furthermore, predictive distribution $p(y_t | y_{1:(t-1)} , x_{1:t})$ is Student's $t$-distribution, hence easily evaluated and simulated from. For details about conjugacy and hyperparameter settings, see the Supplementary Materials (Section~\ref{app:DLM}).

In the DLM synthesis, $\Phi_t = \theta_t$ is the calibration parameter ($\nu_t$ is marginalized out). The conjugacy of the DLMs enables the analytical computation of, and the direct simulation from, the online posterior and prior distributions, $p(\Phi _t | x_{1:t} , y_{1:t})$ and $p(\Phi_{t+1} | x_{1:t} , y_{1:t})$, and the predictive distribution, $p(y_{t+1}| x_{1:t+1} , y_{1:t})$. Utilizing this analytical property, the SMC method for posterior computation can be customized accordingly, as discussed in Secion~\ref{subsec:rao}.

In what follows, the discount factors, $\beta$ and $\delta$, are assumed to be fixed. To calibrate the discount factors, one needs to evaluate the posterior distribution of the calibration parameters multiple times for various values of the discount factors, where the fast computation by the SMC method becomes integral. We will come back to this problem in Section~\ref{sec:discount}

\subsubsection*{Other synthesis functions}
Models in the literature of forecast density combination can be viewed as the synthesis functions in the DBPS framework. For example, \cite{billio2013time} proposed the SMC method for the time-varying weight (TVW) models with the corresponding synthesis function of the form,
\begin{equation*}
    \alpha(y_t | x_t , w_t) = N(y_t | w_t ^\prime x_t , \sigma^2),
\end{equation*}
where $\sigma^2>0$ is constant to be fixed or estimated (and could be replaced with the stochastic volatility model). The weight parameters, $w_t=(w_{1t},\dots ,w_{Kt})'$, are restricted to the simplex ($w_{kt} \geq 0$ for $k \in 1:K$ and $\sum_k w_{kt} = 1$). 
While this constraint on $w_t$ helps interpretation in the context of model averaging, it is not a requirement from the original BPS theory and could be a hindrance to adaptive forecasting and computation. In terms of model flexibility, the non-negativity and sum-to-unity constraints, as well as the lack of the intercept term ($\theta_{0t}$ in the DLM synthesis), could limit the ability of bias adjustment in prediction. In computation, unlike the DLM synthesis, the online posterior and prior under the TVW synthesis are not tractable due to the nonlinearity in the state variables, which limits the computational method applicable to this synthesis model. In fact, \cite{billio2013time} used the vanilla SMC for the online analysis of the TVW synthesis (or the bootstrap filter in Section~\ref{subsec:vanilla}). The comparative analysis of the DLM and TVW syntheses has been made in \cite{mcalinn2019dynamic}, which shows the flexibility of the DLM synthesis improves predictive performance in empirical studies. 

The approach using the TVW synthesis has advanced by sophisticating the synthesis function. \cite{casarin2023flexible} uses the additive noises with stochastic volatilities for each agent forecast before taking their weighted average. The sophisticated synthesis function requires more effort for computation; the bootstrap filter for this model has been improved by M-filtering \citep{bacsturk2019forecast}. Our study differs from the series of research on the TVW synthesis in using the DLM synthesis--- a simple, linear and Gaussian model without any hard constraint--- and in fully utilizing its analytical tractability in devising an efficient SMC algorithm.

\section{Sequential Monte Carlo methods} \label{sec:smc}
To obtain the final output, $p(y_{t+1} | \mathcal{H}_{1:(t+1)} , y_{1:t})$, we have to compute the online posterior distribution $p(\Phi_{1:t} , x_{1:t} | \mathcal{H}_{1:t} , y_{1:t})$. 
We first describe the vanilla SMC method generally applicable to DBPS in the style of sequential importance resampling \citep{rubin1988using}. Then, we provide a custom SMC method for the DLM synthesis. We also comment on the MCMC method for DBPS and its use as a remedy for the problem of particle degeneracy. 

\subsection{SMC for DBPS} \label{subsec:vanilla}
Denote the pair of calibration parameters and agent forecasts by $Z_t = (\Phi_{1:t},x_{1:t})$. The goal of the SMC method is to update the online posteriors from $p(Z_{t-1} | \mathcal{H}_{1:(t-1)} , y_{1:(t-1)})$ to $p(Z_t | \mathcal{H}_{1:t} , y_{1:t})$. Suppose that, at time $t{-}1$, we have particles $\{ Z_{t-1}^i , W_{t-1}^i \}_{i=1:M}$ to approximate the online posterior $p(Z_{t-1} | \mathcal{H}_{1:(t-1)} , y_{1:(t-1)} )$ by 
\begin{equation*}
    \hat{p}^M_{t-1} (Z_{t-1} ) = \sum _{i = 1}^M W_{t-1}^i \delta _{Z_{t-1} ^i} (Z_{t-1}),
\end{equation*} 
where $\{ W_{t-1}^i \}$ are weights and satisfy $W_{t-1}^i \geq 0 $ for $i = 1:M$ and $\sum_i W_{t-1}^i = 1$. 
Then, our objective is to obtain a particle approximation of the online posterior distribution at the next time point, or $p(Z_t | \mathcal{H}_{1:t} , y_{1:t})$. 

In DBPS, $x_t$ is independent of $(Z_{t-1} , \Phi_t)$ conditional on $( y_{1:(t-1)} , \mathcal{H}_{1:t} )$. Thus, we have 
\begin{equation*}
\begin{split}
    p(Z_t | \mathcal{H}_{1:t} , y_{1:t}) &\propto p(y_t | Z_t , \mathcal{H}_{1:t} , y_{1:(t-1)}) p(Z_t | \mathcal{H}_{1:t} , y_{1:(t-1)}) \\
    &=  \alpha(y_t |x_t, \Phi_t)  p(x_t , \Phi_t | Z_{t-1}, \mathcal{H}_{1:t} , y_{1:(t-1)})p( Z_{t-1}| \mathcal{H}_{1:t} , y_{1:(t-1)})\\
    &=  \alpha(y_t |x_t, \Phi_t)  p(\Phi_t | \Phi_{1:(t-1)}, x_{1:(t-1)}, y_{1:(t-1)})h_t(x_t) p( Z_{t-1}| \mathcal{H}_{1:(t-1)} , y_{1:(t-1)}).
\end{split}
\end{equation*}
Here, $p(\Phi_t | \Phi_{1:(t-1)}, x_{1:(t-1)}, y_{1:(t-1)})$ is dependent on the choice of synthesis function $\alpha$, and online posterior $p( Z_{t-1}| \mathcal{H}_{1:(t-1)} , y_{1:(t-1)})$ is approximated by the aforementioned particles. 
In approximating the online posterior at time $t$, we use proposal distribution $q_t(x_t , \Phi_t)$ as
\begin{equation*}
    p(Z_{t} | \mathcal{H}_{1:t} , y_{1:t} ) \propto \alpha(y_t |x_t, \Phi_t)  \frac{p(\Phi_t | \Phi_{1:(t-1)}, x_{1:(t-1)}, y_{1:(t-1)})h_t(x_t)}{q_t(x_t , \Phi_t)} q_t(x_t , \Phi_t) p( Z_{t-1}| \mathcal{H}_{1:(t-1)} , y_{1:(t-1)}). 
\end{equation*}
Based on this expression, we can construct the particle approximation as follows. First, we sample $Z_{t-1}^i$'s from $\hat{p}^M_{t-1} (Z_{t-1})$, by resampling $Z_{t-1}^{1:M}$ with probability $W_{t-1}^{1:M}$. Next, we generate $(x_t^i , \Phi_t^i)$ from $q_t(x_t , \Phi_t)$ and calculate the weight $W_t ^i$ for particle $Z_t^i = (Z_{t-1}^i , x_t^i , \Phi_t ^i)$ by
\begin{equation*}
     W_t^i = \frac{w_t^i}{\sum_{j = 1}^M w_t^j} \ \ \ \mathrm{and} \ \ \ w_t ^i = \alpha(y_t |x_t^i, \Phi_t^i)  \frac{p(\Phi_t^i |Z_{t-1}^i, y_{1:(t-1)})h_t(x_t^i)}{q_t(x_t^i , \Phi_t^i)}  \ \ \text{for } i \in 1:M.
\end{equation*}
Finally, the target distribution, $p(Z_t | \mathcal{H}_{1:t},y_{1:t})$, can be approximated by
\begin{equation*}
        \hat{p}_t^M (Z_t) = \sum _{i = 1}^M W_{t}^i \delta _{Z_t ^i} (Z_t).
\end{equation*}
In using this algorithm, the choice of proposal distribution $q_t(x_t , \Phi_t)$ is the key. Bootstrap particle filter (BPF, \citealt{gordon1993novel}) uses $q_t(x_t , \Phi_t) = h_t(x_t)p(\Phi_t | \Phi_{1:(t-1)}, x_{1:(t-1)}, y_{1:(t-1)})$ and is applicable to a wide class of synthesis functions. \cite{billio2013time} uses BPF for the posterior computation under the TVW model, where $p(\Phi_t | \Phi_{1:(t-1)}, x_{1:(t-1)}, y_{1:(t-1)})$ is simply a Gaussian random walk and easy to simulate from. There is a series of research on a better choice of $q_t(x_t , \Phi_t)$ including, for example, the auxiliary particle filter \citep{pitt1999filtering} and the resample-move strategy \citep{gilks2001following}. In the DLM synthesis, we can utilize the analytical property of the model and derive the custom proposal distribution, as we will see below.

\subsection{Rao-Blackwellization for DLM synthesis} \label{subsec:rao}
In the previous subsection, we approximate the {\it joint} online posterior $p(Z_{t-1} | \mathcal{H}_{1:(t-1)} , y_{1:(t-1)})$ to compute the online prior $p(\Phi_t | \mathcal{H}_{1:(t-1)} , y_{1:(t-1)})$. In some cases, this distribution can be evaluated without using the joint online posterior as 
\begin{equation*}
    p(\Phi_t | \mathcal{H}_{1:(t-1)} , y_{1:(t-1)}) = \int p(\Phi_t | x_{1:(t-1)} , \mathcal{H}_{1:(t-1)} , y_{1:(t-1)})p(x_{1:(t-1)} | \mathcal{H}_{1:(t-1)} , y_{1:(t-1)}) dx_{1:(t-1)}    .
\end{equation*}
Thus, if the {\it marginal} online posterior $p(x_{1:(t-1)} | \mathcal{H}_{1:(t-1)} , y_{1:(t-1)})$ is available, we do not need the joint online posterior $p(Z_{1:(t-1)} | \mathcal{H}_{1:(t-1)} , y_{1:(t-1)})$. By marginalizing out $\Phi_{1:t}$, we can improve not only the efficiency of the algorithm but also the computational time. This approach is known as the Rao-Blackwellized particle filter \citep{doucet2000sequential}. 

Suppose that we have particles $x_{t-1}^{1:M}$ and weights $W_{t-1}^{1:M}$ to approximate $p(x_{1:(t-1)} | \mathcal{H}_{1:(t-1)} , y_{1:(t-1)})$ by $\hat{p}^M_{t-1}(x_{1:(t-1)})$. With the proposal distribution $q_t(x_t)$, the online posterior distribution of $x_{1:t}$ can be written as 
\begin{equation*}
\begin{split}
    p(x_{1:t} | \mathcal{H}_{1:t} , y_{1:t}) &\propto p(y_t | x_{1:t} , y_{1:(t-1)}) \frac{h_t(x_t)}{q_t(x_t)} q_t(x_t) p(x_{1:(t-1)} | \mathcal{H}_{1:(t-1)}, y_{1:(t-1)}).
    \end{split}
\end{equation*}
Then, generate $x_t^i \sim q_t(x_t)$ for each $i=1:M$, so that we can approximate the marginal online posterior, $p(x_{1:t} | \mathcal{H}_{1:t} , y_{1:t})$, by
\begin{equation*}
    \hat{p}^M(x_{1:t}) = \sum _{i = 1}^M W_t^i \delta_{x_{1:t}^i}(x_{1:t}), 
\end{equation*}
where the weights are computed as 
\begin{equation*}
     W_t^i = \frac{w_t ^i}{\sum_{j = 1}^M w_t^j} \ \ \ \mathrm{and} \ \ \ w_t^i = p(y_t | x_{1:t}^i , y_{1:(t-1)}) \frac{h_t(x_t^i)}{q_t(x_t^i)} \ \ \text{for } \ i \in 1:M.
\end{equation*}
For the Rao-Blackwellized particle filter to be feasible, density $p(y_t | x_{1:t}^i , y_{1:(t-1)})$ used in weight $w_t^i$ must be evaluated fast.

In the case of the DLM synthesis, the weights can be obtained in closed form and easily evaluated. As pointed out in Section~\ref{subsec:DBPS}, the density of interest, $p(y_t | x_{1:t}^i , y_{1:(t-1)})$, is the Student's $t$-distribution; for details about the functional form, see Section~\ref{app:DLM}. 
Then, we use $q_t(x_t)=h_t(x_t)$ as the proposal distribution, so that the weight is computed simply by $w_t^i = p(y_t | x_{1:t}^i, y_{1:(t-1)})$. 
Therefore, for the DLM synthesis, we only have to draw $x_t$ from $h_t (x_t)$ and set weights $w_t^i = p(y_t | x_{1:t}^i , y_{1:(t-1)})$.

\subsection{Interventions by MCMC}\label{subsec:MCMC}
\subsubsection{Gibbs sampler}
The Gibbs sampler has been the standard method to compute the online posteriors in the context of DBPS. The algorithm tailored for the DLM synthesis is provided in \cite{mcalinn2019dynamic} and summarized as follows. 
To estimate the posterior distribution at time $T$, $p(x_{1:T},\Phi_{1:T} | y_{1:T},\mathcal{H}_{1:T})$, we iteratively sample particles from the following full conditionals: 
\begin{equation*}
\begin{split}
    \Phi_{1:T}^n | x_{1:T}^{n-1}, y_{1:T}  &\sim p(\Phi_{1:T} | x_{1:T}^{n-1} , y_{1:T}) \\
    x_{1:T}^n | \Phi _{1:T}^{n}, y_{1:T}  &\sim \prod _{t = 1}^T \alpha(y_t | x_t , \Phi_t ^{n}) h_t(x_t),
\end{split}
\end{equation*}
for $n \in 1:N$, where $N$ is the chain size of MCMC. 
In this study, we use all the historical data, $y_{1:T}$, in implementing the MCMC method. For large datasets, one could instead use the latest $T_0$ observations, $y_{T-T_0-1:T}$, for computational feasibility (Practical filtering; e.g., \citealt{prado2021time}). 

In the DLM synthesis, sampling $\Phi_{1:T}^n$ from $p(\Phi_{1:T} | x_{1:T}^{n-1} , y_{1:T})$ can be implemented easily by the FFBS algorithm (\citealt{fruhwirth1994data,carter1994gibbs}). In sampling $x_{1:T}^n$, we utilize the fact that all agent forecasts are Student's $t$-distributions and expressed as the scale mixture of normal distributions; $h_{kt}(x_{kt}) = \int N(x_{kt} | \mu_{kt} , \sigma_{kt}^2 H_{kt})p(\sigma_{kt} ^2)d\sigma_{kt}^2$, where $\sigma _{kt}^2$ follows some inverse-gamma distribution. Conditional on latent scale $\sigma _{kt}^2$, the posterior distribution of $x_{t}$ is simply a normal distribution. 
Thus, we simulate latent scale $\sigma_{kt}^2$ together as follows: 
\begin{equation*}
    \begin{split}
        \Phi_{1:T}^n | x_{1:T}^{n-1}, y_{1:T} &\sim  p(\Phi_{1:T} | x_{1:T} , y_{1:T})\ \  \text{by FFBS} \\
        x_{t}^n | \Phi_{1:T}^n, (\sigma_{1:T}^2)^{n-1}, y_{1:T} &\sim \text{Normal} \ \ \text{for} \ t = 1:T \\
        (\sigma_{kt}^2)^n | \Phi_{1:T}^n, x_{1:T}^{n}  &\sim \text{Inverse-Gamma} \ \ \text{for} \ k = 1:K, t = 1:T.
    \end{split}
\end{equation*}
Note that $x_t$'s and $\sigma_{kt}^2$'s can be sampled in parallel. For details, see Section~\ref{app:DLM}.

\subsubsection{MCMC intervention}
Particle degeneracy is an inevitable problem in using the SMC methods repeatedly. As $t$ increases, it is likely to have fewer particle variations, or increase the variance of the weights, resulting in an inaccurate approximation of the online posteriors. To avoid this problem, it is advised that one should monitor an efficiency measure and, once the measure gets below some pre-specified threshold, then intervene in the algorithm to recover the accuracy of the approximation (e.g.,~\citealt{gilks2001following,chopin2002sequential}). 
We take the effective sample size (ESS) as the monitoring measure which is computed by using the particles generated by the SMC method as  
\begin{equation*}
    \text{ESS}_t = \frac{1}{ \sum _{m = 1}^M (W_t ^m)^2  }, 
\end{equation*}
We set some threshold $C>0$ and, if $\text{ESS}_t < C$, we discard all the particles obtained by the SMC method at time $t$ and generate a new set of particles using the MCMC method. 
That is, we re-do the approximation of the online posterior at time $t$ by the Gibbs sampler with chain size $N$. Then, using the generated chain $\{ x_{1:t}^{1} , \cdots ,  x_{1:t}^{N}  \}$ (and ignoring the chains of $\Phi_t$'s), we approximate the online posterior by 
\begin{equation*}
    \hat{p}_t^N( x_{1:t}) = \sum_{n = 1}^N \frac{1}{N} \delta_{x_{1:t}^{n}}(x_{1:t}).
\end{equation*}
This distribution obtained by the MCMC method converges weakly to $p(x_{1:T}|y_{1:T})$ as in the SMC method, so we use it as the approximation of the online posterior at time $t$. At the next time point of $t+1$, we implement the SMC method by using $\hat{p}_t^N( x_{1:t})$. Thus, the proposed algorithm is viewed as the combination of the Rao-Blackwellized particle filter and the Gibbs sampler, as summarized in Algorithm~\ref{alg:SMCwithMCMC}.  

Note that the chain size $N$ of the MCMC method can differ from the particle size $M$ of the SMC method. In our application, we set $N=M$ for simplicity. Since the MCMC method is efficient but time-consuming, one can set $N<M$ in practice. The threshold $C$ should be chosen carefully. The larger $C$ is, the more interventions are expected, requiring longer computational time. It is essential to keep $C$ moderate so that the sequential computation is feasible while the accuracy of the particle approximation remains satisfactory. We conducted the sensitivity analysis on the choice of $C$ in the real data application, reported in Section~\ref{app:BPS}.

\begin{figure}[!t]
\begin{algorithm}[H]
    \caption{Rao-Blackwellized Bootstrap Particle Filter with MCMC intervention}
    \label{alg:SMCwithMCMC}
    \begin{algorithmic}    
    \FOR{$m = 1:M$}
        \STATE Generate $x_1^{m} \sim h_1(x_1)$ 
        \STATE Set $w_1^m = p(y_1 | x_1^m) $ 
        \ENDFOR 
    \STATE Set $W_1^m = w_1^m / \sum_i w_1^i$ for $m = 1:M$
    \STATE Set $\hat{p}(x_{1}) = \sum_{i = 1}^M W_1^i \delta_{x_1^i}(x_1)$
    \FOR{$t = 2:T$}
        \FOR{$m = 1:M$}
            \STATE Generate $x_{1:(t-1)}^m \sim \hat{p}(x_{1:(t-1)})$
            \STATE Generate $x_t^{m} \sim h_t(x_t)$ 
            \STATE Set $w_t^m = p(y_t | x_{1:t}^m , y_{1:(t-1)}) $ 
            \ENDFOR 
        \STATE Set $W_t^m = w_t^m / \sum_i w_t^i$ for $m = 1:M$
        \STATE Set $\hat{p}(x_{1:t}) = \sum_{i = 1}^M W_t^i \delta_{x_{1:t}^i}(x_{1:t})$
        \IF{$\text{ESS}_t < C$}
        \STATE Generate a chain $\{x_{1:t}^1 , \cdots , x_{1:t}^N \}$ from MCMC with chain size $N$.
        \STATE Set $\hat{p}(x_{1:t}) = \sum_{i = 1}^N \delta_{x_{1:t}^i}(x_{1:t})/N$
        \ENDIF
        \ENDFOR
    \end{algorithmic}
\end{algorithm}
\end{figure}

\section{Numerical evaluation of the proposed SMC methods}\label{sec:data}
In this section, we apply the proposed SMC method to the predictions of the quarterly US inflation rates. This numerical study aims to confirm that the predictive performance of the SMC method is as competitive as the MCMC method with a sufficiently large chain size. Other properties of the SMC method, such as the necessity of occasional MCMC interventions, are also illustrated in this example. The settings of this numerical example, including the dataset and agent models, have been used in \cite{mcalinn2019dynamic} to evaluate the predictive performance of the DBPS.

\subsection{Settings} \label{subsec:setting}

\subsubsection*{Data description}
We predict the annual percentage change in a chain-weighted GDP price index $(y_t)$ as an inflation index. 
We also use the unemployment rate $(u_t)$ and the yield on three-month Treasury bills as short-term nominal interest rate $(r_t)$ as covariates. We use the data observed from 1961/Q1 to 2022/Q4 ($t = 1:248$). The data is retrieved at the webpage of the Federal Reserve Bank of St.~Louis (\url{https://research.stlouisfed.org/}).

Our numerical study is largely based on \cite{mcalinn2019dynamic}; we decided to stick to the univariate data and simple agent models to focus on the computational problem of our main interest. 
One notable change from the experiment in \cite{mcalinn2019dynamic} is the use of recent data, especially those observed in 2020-2022. In this period, when the COVID-19 pandemic hit the US and world economies, a sudden burst of the inflation rate was observed, as seen later in Figure~\ref{fig:pred_BPS} (For unemployment and interest rates, see Figure~\ref{fig:data}). Sequential inference and prediction by the SMC method became challenging in this period, as discussed below.

\subsubsection*{Agent models}

We use four agent models ($K=4$) to create predictive distributions $\mathcal{H}_t$. All agents are DLMs but with different predictors: $\mathcal{M}_1$ - $y_{t-1}$ ; $\mathcal{M}_2$ - $y_{t-3:t-1} , u_{t-3:t-1} , r_{t-3:t-1}$ ; $\mathcal{M}_3$ - $y_{t-3:t-1}$  ; $\mathcal{M}_4$ - $y_{t-1} , u_{t-1} , r_{t-1}$. 
We use the same hyperparameters for the four agent models: $(m_0 , C_0 , n_0 , s_0 , \beta , \delta ) = (\mathbf{0} , \mathbf{I} , 2 , 0.01 , 0.99 , 0.95)$, where $\mathbf{I}$ is the identity matrix. At each $t$, we learn about these four models from the observed data to obtain the set of predictive densities, $\mathcal{H}_t = \{ h_{1t} , \cdots , h_{4t}\}$, where $h_{kt}(y_t) = p(y_t | y_{1:t-1} , u_{1:t-1} , r_{1:t-1} , \mathcal{M}_k)$. Since the agent models are all DLMs, the predictive densities, $h_{kt}(\cdot )$'s, are simply the Student's $t$-distributions and can be computed without any simulation-based method. 

\subsubsection*{DBPS}

We use the DLM synthesis in Section~\ref{subsec:DBPS} with hyperparameters $(m_0 , C_0 , n_0 , s_0 , \beta , \delta )= ( (0 , \mathbf{1}^\prime/4)^\prime , \mathbf{I} ,$ $ 10 , 0.002 , 0.99 , 0.95)$ where $\mathbf{1}$ is the 4-dimensional vector with all elements $1$. 

\subsubsection*{Training and test datasets}

We divide the whole dataset into the three sub-datasets: the first learning period from 1961/Q1 to 1977/Q1 $(t = 1:65)$, the second learning period from 1977/Q2 to 1989/Q4 $(t = 66:116)$, and the evaluation period from 1990/Q1 to 2022/Q4 $(t = 117:248)$. In the first learning period, we only learn about the agent models and predictive distributions. In the second learning period, we keep learning about the agent models but start to estimate the calibration parameters in the DLM synthesis. However, we do not evaluate its predictive performance yet. In the evaluation period, we still keep model learning and calibration, and we assess the synthesized predictive distribution $p(y_t | y_{66:t-1} , \mathcal{H}_{66:t})$ at each time $t$, computing the point and intervals estimations and other measures introduced below.

\subsubsection*{Computational details}

For comparative analysis, we take three approaches to the sequential analysis of the inflation data. First, we implement the Rao-Blackwellized particle filter with MCMC interventions. We set the SMC particle size $M = 10000$, the intervention threshold $C = M/20 = 500$, and the MCMC chain size $N = M = 10000$. 
Second, we implement the same SMC method without MCMC interventions. In this fully SMC method, we set the particle size $M = 100000$. As we see later, the problem of particle degeneracy is unavoidable even with this large particle size. Third, we employ the Gibbs sampler repeatedly at every $t$ as practiced in the literature. Depending on the purpose of the analysis, we use different values for chain size $N$.

All the computations are implemented in \textsf{R} on a laptop  computer with Intel Core i7-7700HQ CPU @2.80GHz, 2.80GHz, RAM-8GB.

\subsection{Computational time and efficiency}

\begin{figure}[htbp]
\centering
\includegraphics[width=15cm]{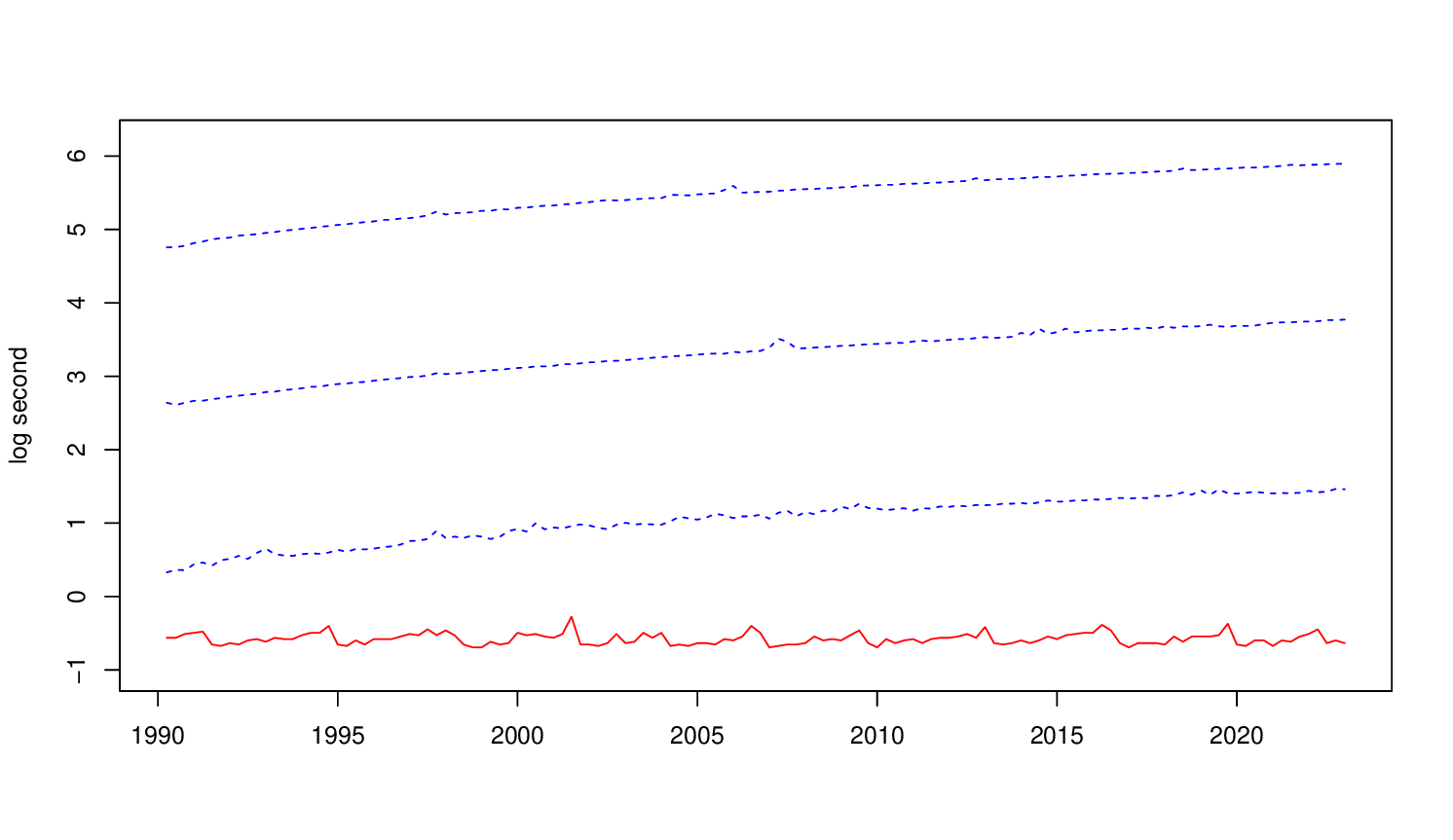}
\caption{Logarithmic computational time (seconds) over the evaluation period by the MCMC method with $N = 100$, $1000$, $10000$ (dotted, from bottom to top) and the SMC method with $M = 10000$ (solid).}
\label{fig:comptime}
\end{figure}

Figure~\ref{fig:comptime} shows the raw computational time (log second) of the proposed SMC method and the repeated use of Gibbs sampler with different chain sizes ($N = 100$, $1000$, and $10000$) at each of $t = 117:T$. Since the Gibbs sampler processes all the historical data and sample $(t-66)$ variables, the computational time increases linearly in $t$. By contrast, the SMC method only processes the data and parameters at time $t$, whose computational time is constant over different time points. The advantage of the SMC method in computation can also be seen in the number of parameters to sample; the Gibbs sampler generates $(x,\Phi,\sigma^2)$, while the SMC method needs to sample $x$ only. For details, see Section~\ref{app:BPS}. The average time of $T-66$ computations by SMC $(M=10000)$ is 0.56 seconds and the maximum time is 0.76 seconds (standard deviation is 0.046). The computational time of the MCMC method $(N=10000)$ at $t = 248$ is 363.5 seconds, while it is 116.36 seconds at $t = 117$. 

\begin{figure}[htbp]
\centering
\includegraphics[width=15cm]{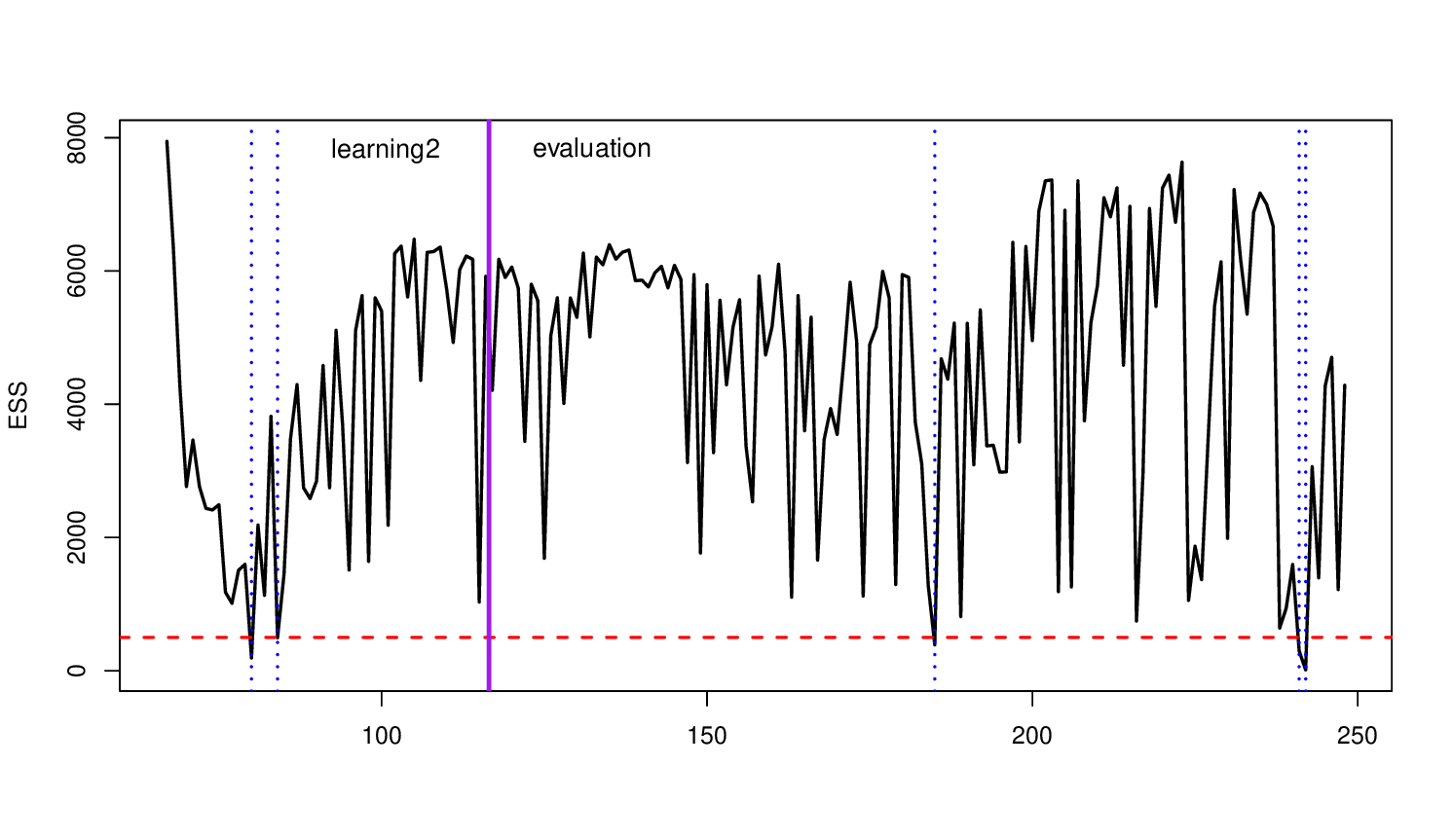}
\caption{Effective sample size (ESS) over $t = 66:248$ (solid). The starting point of the evaluation period ($t=117$) is indicated by the vertical solid line. The five MCMC interventions with chain size $N = 10000$ (time points when $\text{ESS}_t < C$) are indicated by the vertical dotted lines. The threshold is set to $C=500$ and indicated by the horizontal dashed line. }
\label{fig:ESSplot}
\end{figure}

In Figure~\ref{fig:ESSplot}, the realized ESS are plotted over $t = 66:248$. The figure also indicates time points when $\text{ESS}_t < C=500$ and the MCMC intervention with size $N=10000$ is triggered. With this threshold, we made the intervention five times out of $183\ (t = 66:248)$ time points. The total time for this computation is 1266.79 seconds, mainly for the five MCMC interventions. Sensitivity analysis about different values of threshold $C$ can be found in Section~\ref{app:BPS}, where the number of interventions, computational time, and predictive performance are reported. 

\subsection{Approximation accuracy}
We assess the accuracy of particle approximation made by the SMC methods for the synthesized predictive distributions, $p(y_t | y_{66:t-1} , \mathcal{H}_{66:t})$, and the posterior distributions of the calibration parameters, $p(\Phi_t | y_{66:t-1} , \mathcal{H}_{66:t-1})$, for $t = 117:248$. 
First, we employ the MCMC method repeatedly, using chain size $N = 70000$ at each time, to obtain the accurate approximation of the posterior and predictive distributions. Then, we implement the computation by the SMC methods with and without the MCMC interventions and compare their results with the repeated MCMC method. 

\begin{figure}[htbp]
\centering
\includegraphics[width=15cm]{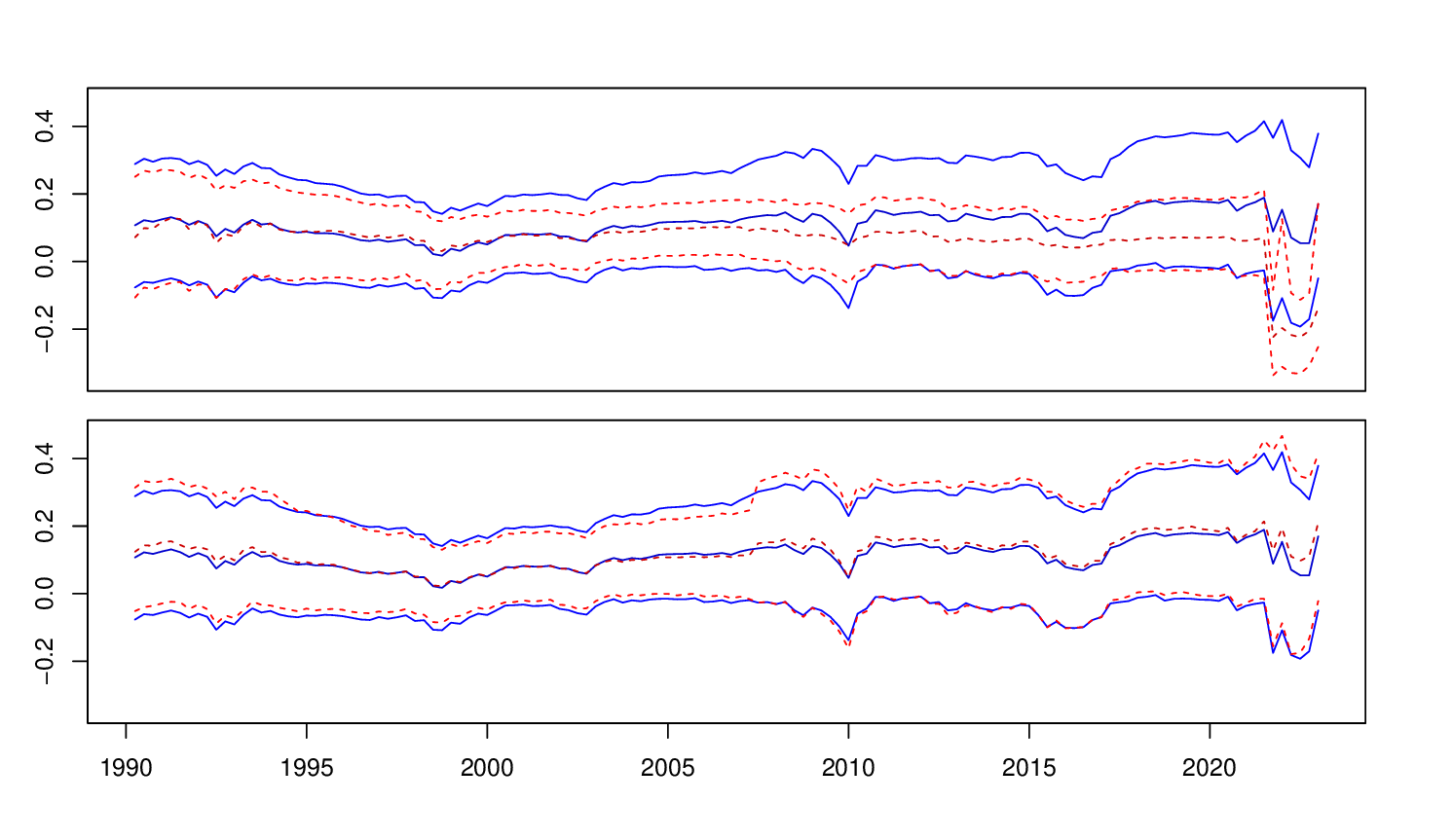}
\caption{The $5\%, 50\%,$ and $95\%$ quantiles of posterior distributions $p(\theta_{t,0}|y_{66:t-1},\mathcal{H}_{66:t-1})$ computed by MCMC (solid) and SMC (dashed). The top panel shows the result of the fully SMC-based method, while the bottom allows the MCMC interventions.}
\label{fig:posteriorplot}
\end{figure}

In Figure~\ref{fig:posteriorplot}, we plot the posterior distributions of the location calibration parameter $\theta_{t,0}$, or $p(\theta_{t,0}|y_{66,t-1},\mathcal{H}_{66,t-1})$, for $t = 117:248$. Without MCMC interventions, the posteriors computed by the generated particles are biased and degenerated. The fully SMC-based approach is also sensitive to the sudden burst of the inflation rate at $t = 242$ (2021/Q2). At this time point, both the agent models and the DLM synthesis significantly fail in prediction, causing a severe particle degeneracy in the sequential analysis. 
With interventions, the SMC method accurately approximates the target predictive distribution; their median and 90\% credible intervals are almost identical to those computed by the MCMC method with the large chain size. We confirmed that the SMC method with interventions can achieve the accurate particle approximation in other posterior plots; see Section~\ref{app:BPS} for details.

\begin{figure}[htbp]
\centering
\includegraphics[width=14cm]{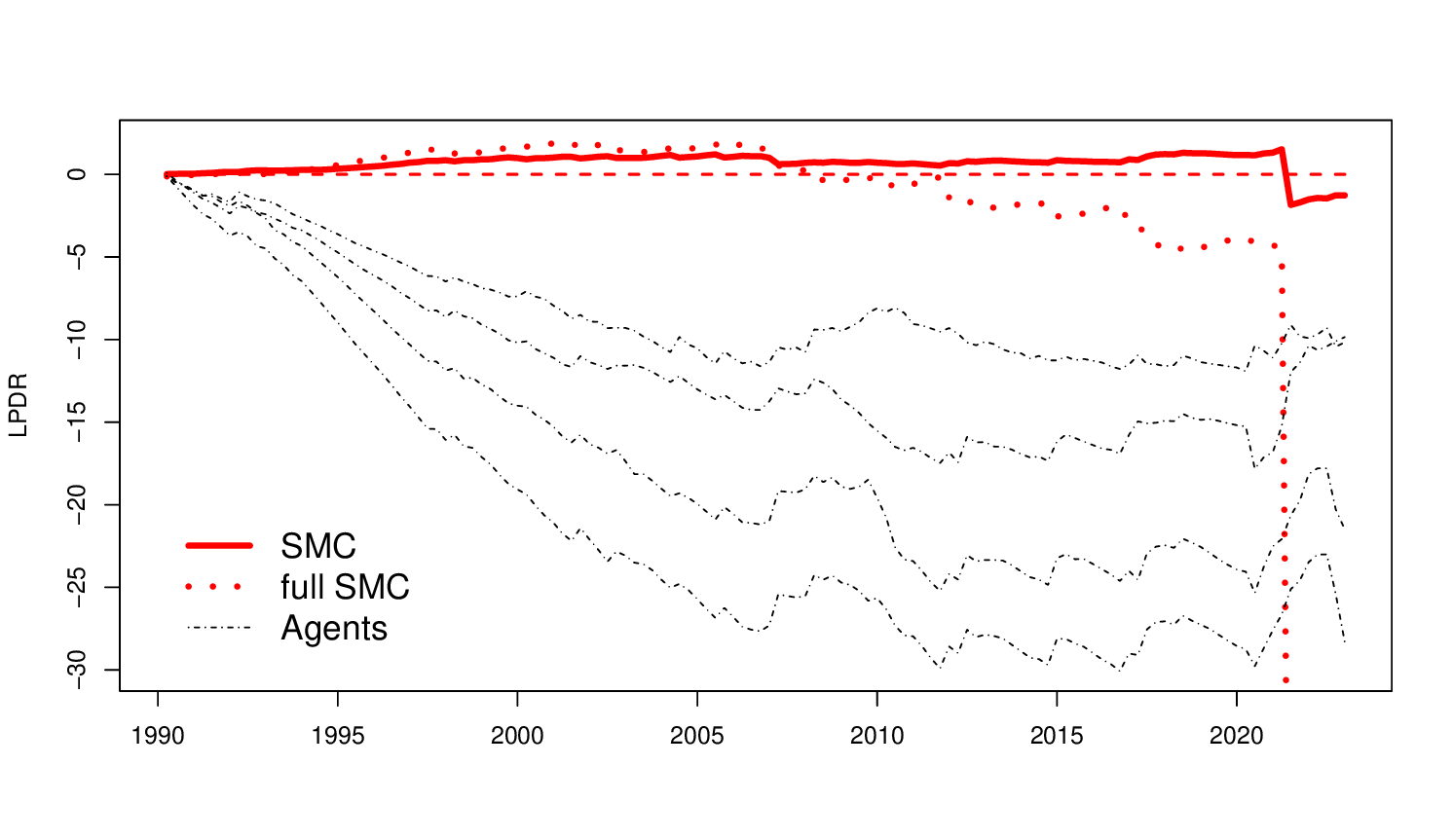}
\caption{Log predictive density rations (LPDRs) of the SMC methods with and without MCMC interventions (solid and dotted, respectively) and four agent models (dash-dotted) against the MCMC method. The LPDR of the SMC method with interventions is close to zero, showing its accuracy of particle approximation. The fully SMC-based approach shows decreasing LPDRs due to the accumulated approximation errors.}
\label{fig:LPDRplot}
\end{figure}

To evaluate the SMC method from the viewpoint of predictive performance, we compute the log predictive density ratios (LPDR) over $t = 117:248$, which is defined by 
\begin{equation*}
    \text{LPDR}_t = \sum_{s = 117}^{t} \log( p(y_s | y_{66:s-1},\mathcal{H}_{66:s}) / p_{\text{MCMC}}(y_s | y_{66:s-1},\mathcal{H}_{66:s}))
\end{equation*}
where $p_{\text{MCMC}}(y_s | y_{66:s-1},\mathcal{H}_{66:s})$ is the predictive density computed by using the MCMC method repeatedly, and $p(y_s | y_{66:s-1},\mathcal{H}_{66:s})$ is the predictive density of interest. If the LPDR is close to $0$, the predictive distribution of interest is as performative as the synthesized predictive distribution computed by the MCMC method. Here, we evaluate the LPDR for the particle approximation of the synthesized predictive density computed by the SMC methods with or without MCMC interventions and the four agent predictive densities. The LPDRs are plotted in Figure~\ref{fig:LPDRplot}. Again, without interventions, the approximation errors accumulate gradually over time and critically at the sudden burst of the inflation rate. The MCMC interventions improve approximation accuracy regarding LPDRs at all time points.

\section{Application to the sequential loss discounting} \label{sec:discount}

In the previous section, we confirmed that the proposed SMC method could accurately approximate the posterior and predictive distributions. However, this does not mean that the synthesized predictive distribution is always useful. Rather, the synthesized predictive distribution performs poorly in 2020-2022, regardless of the computational methods. This is all due to the sudden burst of the inflation rate, to which all four agent models are slow to adapt. It is also true for the calibration parameters in the DLM synthesis that adapting to this sudden change is difficult, due to the use of the Gaussian random walk as the prior. 

To make the predictive analysis more flexible, we should not fix the discount factors, $\beta$ and $\delta$, in the DLM synthesis, but adaptively change their values. In doing so, one can take a decision-theoretic approach and make a manual intervention to the sequential analysis (\citealt[Chapter~11]{West1997}). Although practiced in applied statistics (e.g., \citealt{chen2018scalable}), this approach involves many subjective decisions, such as deciding the criterion for interventions. Using priors for discount factors is a formal Bayesian approach (e.g.,~\citealt{irie2022sequential}), but involves more intensive use of the SMC methods (e.g.,~\citealt{liu2001combined}). As an alternative easily implemented with fewer manual interventions, one can calibrate discount factors--- trying different values of discount factors and comparing the resulting synthesized predictive distributions. This approach requires the computation of the synthesized predictive distribution multiple times, for which fast computation by the proposed SMC method is crucial.

\subsection{Loss Discounting Framework}
In the presence of possible structural changes, power discounting has been utilized in combining the predictive densities (e.g., \citealt{zhao2016dynamic}). This idea is formalized as the loss discounting framework (LDF) in the context of forecast combination \citep{bernaciak2022loss}. Suppose that we have $K$ agent models and their predictive densities, $h_{kt}(y_t) = p(y_t | y_{1:t-1} , \mathcal{M}_k)$, for $k = 1:K$. In evaluating the historical performance of the agent models, the LDF uses log-discounted predictive likelihood (LDPL) defined by 
\begin{equation*}
    \text{LDPL}_{k,t}(\gamma_1) = \sum_{s = 1}^t \gamma_1 ^{(t - s)} \log ( p(y_s | y_{1:s-1} , \mathcal{M}_k) ) ,
\end{equation*}
where $\gamma _1 \in (0,1]$ is called the first layer discount factor. If $\gamma _1 = 1$, then the LDPL reduces to the log-marginal likelihood of agent model $\mathcal{M}_k$, which is used in the standard Bayesian analysis. The LDPL rates the recent predictive performance of the agent model higher than those in the past, making the model evaluation sensitive to sudden structural changes. Using the LDPLs, the $K$ predictive densities are averaged as 
\begin{equation*}
\begin{split}
    p_{\mathrm{LDF}}(y_t | y_{1:t-1}, \mathcal{H}_{1:t} , \gamma _1) = \sum_{k = 1}^K w_{k}( \text{LDPL}_{1:K,t-1}(\gamma_1))   p(y_t | y_{1:t-1} , \mathcal{M}_k)  ,
\end{split}
\end{equation*}
where the weight function, $w_{k}(\cdot)$, depends on the LDPLs. If $w_{k}$ is the expit function, the LDF reduces to the dynamic model averaging \citep{raftery2010online} and, in the case of $\gamma_1=1$, it is the Bayesian model averaging. 
\cite{bernaciak2022loss} proposes two weight functions: softmax and argmax weights defined by 
\begin{equation*}
\begin{split}
    \text{softmax-weights} (a_1 , \cdots, a_K) &= \left(\frac{\exp(a_1)}{\sum_k \exp(a_k)} , \cdots, \frac{\exp(a_K)}{\sum_k \exp(a_k)} \right), \\
    \text{argmax-weights} (a_1 , \cdots, a_K) &= \Big( 0 , \cdots , \underbrace{1}_{k_{\max}\text{-th}} , \cdots , 0 \Big) , \ \ \ \ \ a_{k_{\max}} \ge a_k, \ \ \ \ \mathrm{for \ all \ }k.
    \end{split}
\end{equation*}
The softmax weights utilize all the $K$ models as in the model averaging, while the argmax weights choose the best model as the final output for prediction. Denote the LDFs with the softmax and argmax weight functions by $\text{LDF}_s$ and $\text{LDF}_a$, respectively. 

The first-layer discount factor, $\gamma_1$, is an important tuning parameter to control how much we discount the past predictive performance of the agent models. To emphasize the dependence on $\gamma_1$, we write the predictive distribution in the LDF as $p_{\mathrm{LDF}_{\ast}} ( 
y_t | y_{1:t-1}, \mathcal{H}_{1:t}, \gamma_1)$, where $\ast \in \{ s,a \}$. To compare different values of $\gamma_1$, we have to obtain multiple predictive distributions with different values of $\gamma_1$. These distributions need to be combined in the LDF again. To be precise, we introduce the second-layer discount factor $\gamma_2\in [0,1]$, evaluate the LDPL of $p_{\mathrm{LDF}_{\ast}} ( 
y_t | y_{1:t-1}, \mathcal{H}_{1:t}, \gamma_1)$ for each value of $\gamma_1$, and combine the predictive densities using a weight function that is not necessarily the same as one used in the first layer combination. We denote this forecast combination by $\mathrm{LDF}_{\ast_1,\ast_2}$, where $\ast_1,\ast_2\in \{ s,a \}$. For example, if we use the softmax weights to combine the predictions made by $\mathrm{LDF}_{a}$, then it is $\mathrm{LDF}_{a,s}$.

Finally, note that we still have to choose the second-layer discount factor, $\gamma_2$ carefully. In this study, we set $\gamma_2$ to a fixed value. Having extra layers is possible in the general LDF, but is expected to have less impact on the final predictive result.

\subsection{LDF for BPS}
The LDF can be implemented quickly and fast, but its flexibility is limited since the resulting density is simply the weighted average of $K$ agent predictive densities. Here, we use the idea of the LDF in calibrating the discount factors of the DLM synthesis function of the DBPS, namely, $\beta$ and $\delta$. 

Denote the set of $J$ possible values of discount factors by $S = \{ (\beta_1,\delta_1) , \dots , (\beta_J,\delta_J) \}$. At each time $t-1$, for each pair $(\beta_j,\delta_j) \in S$, we compute the synthesized predictive distribution, $p(y_t | \mathcal{H}_{1:t} , y_{1:(t-1)}, \beta _j, \delta_j)$. Consequently, we have $J$ synthesized predictive distributions, which are combined by using the LDF discount factor, $\gamma$, and the softmax or argmax weight function with argument $\text{LDPL}_{j,t}(\gamma)$ computed by $p(y_t | \mathcal{H}_{1:t} , y_{1:(t-1)}, \beta _j, \delta_j)$. We denote this approach by $\text{LDF}_{B,\ast}$, where $\ast \in \{ a,s \}$. From the viewpoint of the two-layer LDF, this approach replaces the first-layer density combination with the DBPS. 

In this approach, we must evaluate the $J$ synthesized predictive distributions at each time point, which could be costly when the computational resource is limited. If these computations are not fully parallelized, then using the MCMC method in a butch process could increase the computational cost, making this approach infeasible. The fast computation by the proposed SMC method is beneficial in this context.

\subsection{Real Data analysis}
Here, we again analyze the US inflation rates using the LDF to calibrate the discount factors. The same agent models and hyperparameters are used, but this time we consider $J=35$ values for the discount factors, $(\beta, \delta)$, in the DLM synthesis function. The 35 synthesized predictive densities are combined in the LDF with discount factor $\gamma = 0.98$. Likewise, we apply the two-layer LDF approach to the same data/agent models with 15 values for $\gamma_1$ in the first layer and $\gamma _2 = 0.98$ in the second layer. For the complete list of the values used for the discount factors, see Section~\ref{app:LDF}. 

In computation, we set particle/chain sizes to $N=M=10000$ and the intervention threshold to $C=700$. The synthesized predictive distributions are computed in the same computational environment described in Section~\ref{subsec:setting}. Note that we cannot fully parallelize the computation of $J=35$ synthesized predictive distributions by using the 8-core CPUs. The raw computational time is about 5 hours in total. For faster computation, one could use fewer values of the discount factor (smaller $J$), especially in using the MCMC method as the intervention. 

\begin{figure}[htbp]
\centering
\includegraphics[width=15cm]{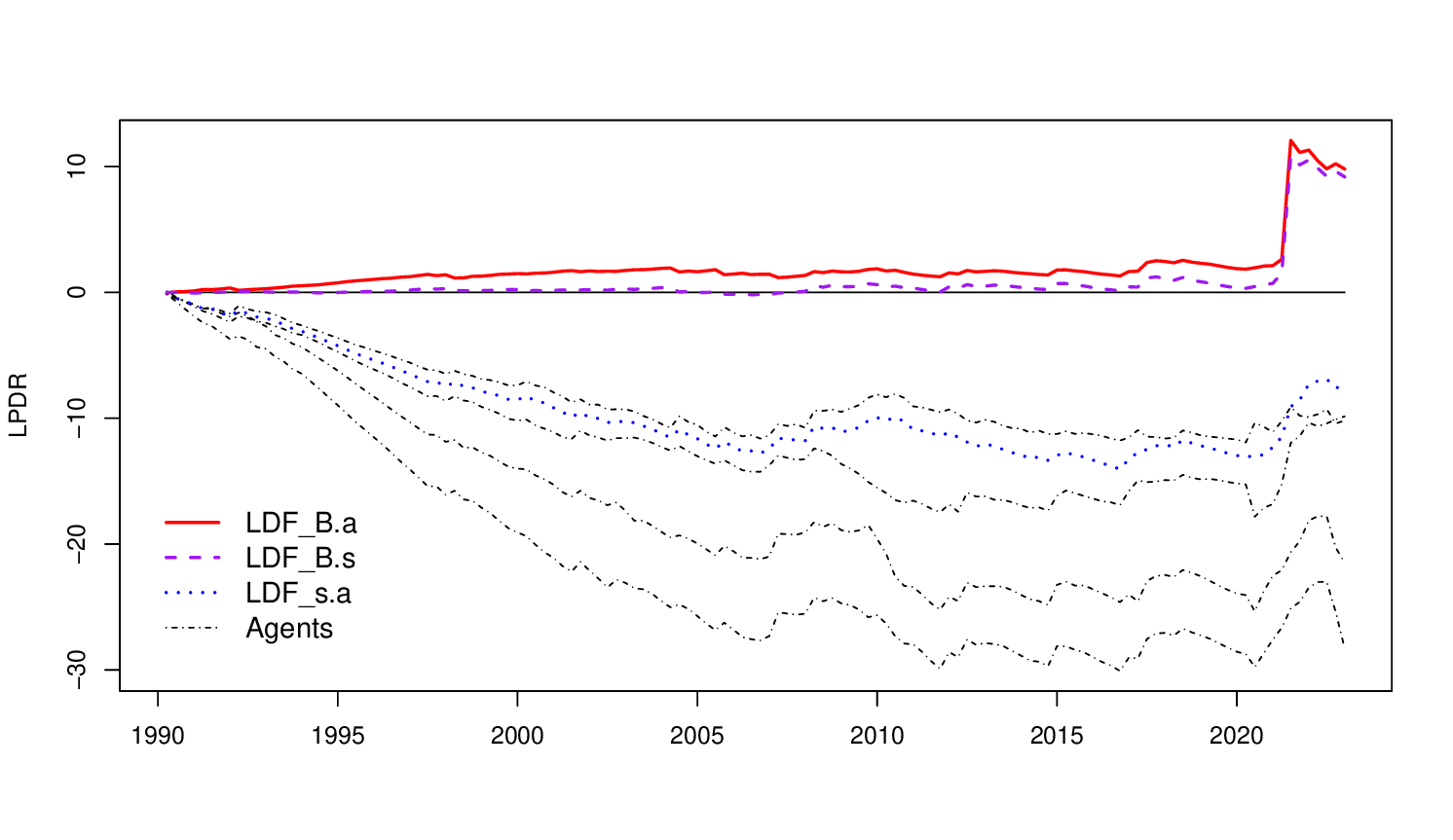}
\caption{LPDRs in the three loss discounting frameworks against the benchmark DBPS with the fixed discount factors: $\text{LDF}_{B,a}$ (solid), $\text{LDF}_{B,s}$ (dashed) and $\text{LDF}_{s,a}$ (dotted). Both $\text{LDF}_{B,a}$ and $\text{LDF}_{B,s}$ improve the prediction by the benchmark DBPS. $\text{LDF}_{s,a}$ is worse than the benchmark, although it shows an improvement of the predictive performance in 2020-2022. }
\label{fig:LDFplot}
\end{figure}

Figure~\ref{fig:LDFplot} shows the LPDRs of the $\text{LDF}_{B,a}$, $\text{LDF}_{B,s}$ and $\text{LDF}_{s,a}$ against the benchmark BPS with fixed discount factors $(\beta , \delta) = (0.99 , 0.95)$. Both the $\text{LDF}_{B,a}$ and $\text{LDF}_{B,s}$ improve the predictive performance of the BPS with the fixed discount factors, especially in the period of the sudden burst in 2020-2022. The $\text{LDF}_{B,a}$ is slightly better than the $\text{LDF}_{B,s}$; in our example, using all of the 35 discount factors might have overestimated the predictive uncertainty. Even with different choices of $\gamma$, the $\text{LDF}_{B,a}$ and $\text{LDF}_{B,s}$ consistently outperform the benchmark DBPS, but an extreme value of $\gamma$, such as $\gamma = 1$, is not recommended. For details, see Section~\ref{app:LDF}.

By contrast, the $\text{LDF}_{s,a}$ is far less performative than the benchmark because the original LDF is simply the weighted average of the agent models and less flexible than the DBPS. However, the advantage of the LDF in discounting the agent predictions in the past can be seen clearly in the increase of its LPDR in 2020-22. Other LDFs provide similar predictions, which are reported in Section~\ref{app:LDF}. 

\begin{figure}[htbp]
\centering
\includegraphics[width=15cm]{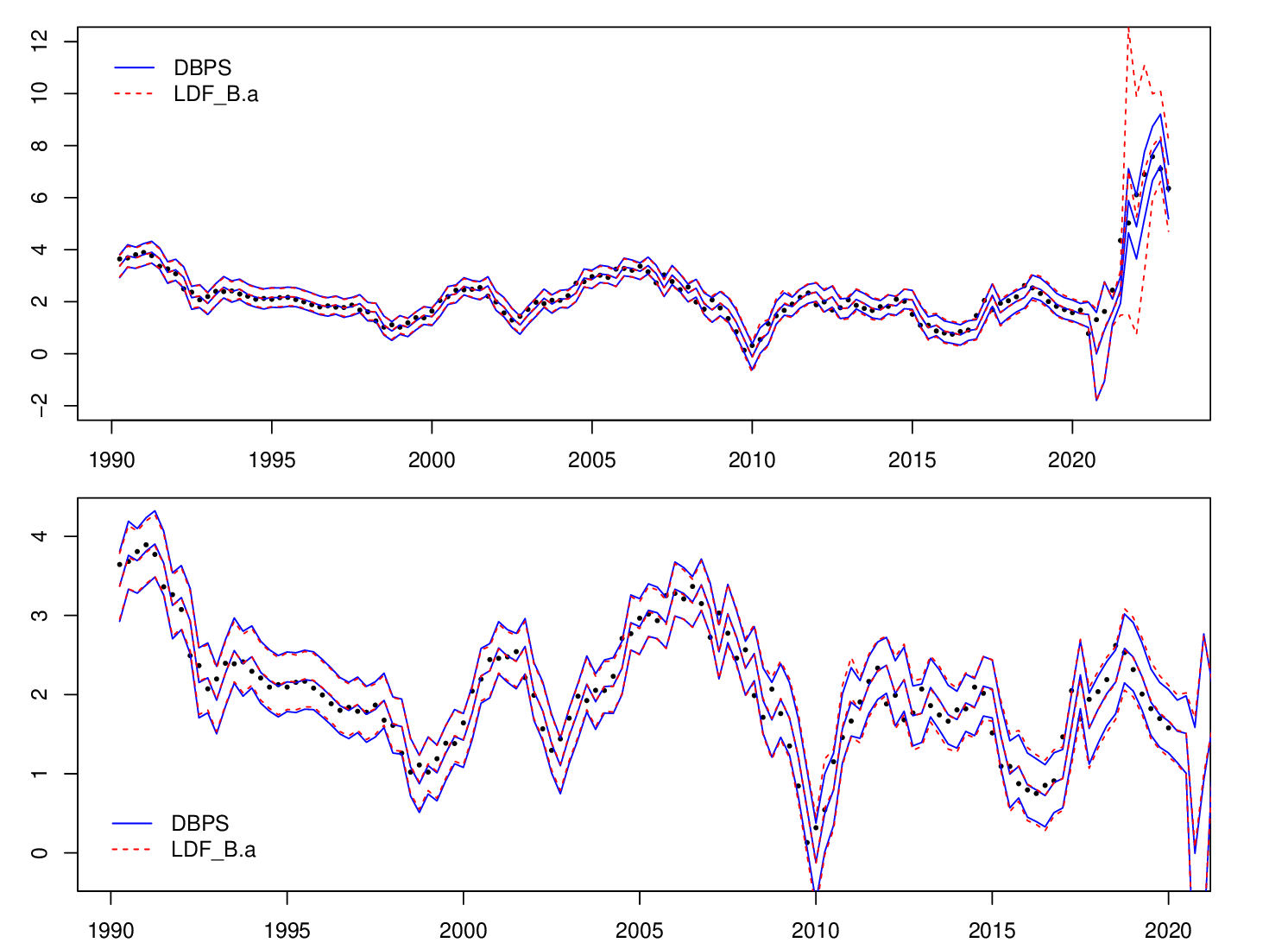}
\caption{Predictive medians and 90\% predictive intervals of DBPS and $\text{LDF}_{B,a}$ with $\gamma = 0.98$ laid over the data in the period $t = 117:248$ (top) and $t = 117:236$ (bottom). $\text{LDF}_{B,a}$ provides diffuse predictive distributions after 2020 due to lower discount factors used for the adaptation to a sudden structural change. The predictive intervals in the $\text{LDF}_{B,a}$ are slightly narrower in 1990-2000 and wider in 2010-2020 than the benchmark DBPS. } \label{fig:pred_BPS} 
\end{figure}

The resulting predictive distributions of the $\text{LDF}_{B,a}$ and benchmark DBPS are shown in Figure~\ref{fig:pred_BPS}. The $\text{LDF}_{B,a}$ benefits from the slightly narrower predictive intervals when the dynamics of the inflation rates are relatively stable (e.g., in 1990-2000) and from the wider intervals when the inflation rates are volatile (e.g., after 2010). 

\section{Future research} \label{sec:future}

The proposed SMC method for DBPS with appropriate interventions by the Gibbs sampler enables the fast computation of the online posterior and synthesized predictive distribution. The fast computation by the SMC method is crucial in not only real-time monitoring but also situations where the online posteriors must be computed multiple times. As an example of the latter cases, we considered the loss discounting DBPS, where the synthesis function has the discount factors to be tuned at each time point. Another important example not studied in this paper is the synthesis of many predictive densities, or the ``large $K$'' problem. In this situation, the synthesized predictive distribution could be less performative since we have many calibration parameters to estimate. In the DLM synthesis, selecting predictive densities in the synthesis function is equivalent to the variable selection problem in the high-dimensional regression. The study of this “model selection” problem in the context of BPS has been hindered due to the computational cost of evaluating the posterior of the calibration parameters, limiting the feasible approaches to using shrinkage priors \citep{chernis2022combining} and clustering \citep{kobayashi2023clustering}. 
The variable selection approach is interesting future research that we believe is enabled by the proposed SMC method. 

The Rao-Blackwellized particle filter proposed in this study is a custom SMC method for the DLM synthesis. 
Although many other SMC methods could be used for sequential forecasting by the DBPS, the problem of particle degeneracy is most likely unavoidable, especially in the period of sudden bursts/structural changes. 
Developing custom computational algorithms for other synthesis functions, such as mixture synthesis \citep{JohnsonWest2022}, is undoubtedly important future research.

\section*{Acknowledgments}
We thank Naoki Awaya for his comments and suggestions. 
The second author's research was partly supported by JSPS KAKENHI Grant Number 22K20132 from Japan Society for the Promotion of Science.

\bibliographystyle{apalike}
\bibliography{bayes}

% --------------------------------------
% --- ONLINE SUPPLEMENTARY MATERIALS ---
% --------------------------------------
\clearpage
\appendix

\setcounter{section}{0}
\renewcommand{\thesection}{S\arabic{section}}
\setcounter{table}{0}
\renewcommand{\thetable}{S\arabic{table}}
\setcounter{figure}{0}
\renewcommand{\thefigure}{S\arabic{figure}}

\begin{center}
{\LARGE {\bf Supplementary Materials for \\``Sequential Bayesian Predictive Synthesis''}}
\end{center}

\section{DLM synthesis}\label{app:DLM}
Here, we describe the details of DLMs, including the forward filter to compute the online posterior and predictive distributions. Note that the DLMs are used as both the agent models and synthesis functions. For further discussions, see \cite{prado2021time}. 

A DLM with discount factors $\beta , \delta \in (0,1]$ is defined as 
\begin{equation*}
\begin{split}
    & y_t = F_t ^\prime \theta_t + \epsilon_t \ , \ \epsilon_t \sim N(0 , \nu_t) \\
    & \theta_t = \theta_{t-1} + \omega_t \ , \ \omega_t \sim N(0 , \nu_t W_t) \\
    & \nu_t = \frac{\beta}{\gamma_t} \ , \ \gamma_t \sim Be \left(\beta \frac{n_{t-1}}{2} , (1-\beta)\frac{n_{t-1}}{2} \right)
\end{split}
\end{equation*}
where $F_t = (1 , x_{1t} , \cdots , x_{Kt})^\prime , \theta_t = (\theta_{0t} , \theta_{1t} \cdots , \theta_{Kt})^\prime,$ and $Be(a,b)$ denotes beta distribution. $W_t$ is defined implicitly so that the following relationship holds;
\begin{equation*}
    \begin{split}
        &p(\theta_{t-1} | \nu_{t-1} , x_{1:(t-1)} , y_{1:(t-1)}) = N \left(\theta_{t-1} \left| m_{t-1} , \frac{\nu_{t-1}}{s_{t-1}} C_{t-1} \right)\right. \\
        & p(\nu_{t-1} | x_{1:(t-1)} , y_{1:(t-1)}) = IG \left(\nu_{t-1} \left| \frac{n_{t-1}}{2} , \frac{n_{t-1} s_{t-1}}{2} \right) \right. \\
        &p(\theta_{t} | \nu_{t} , x_{1:(t-1)} , y_{1:(t-1)}) = N \left(\theta_{t} \left| a_{t} , \frac{\nu_{t}}{s_{t-1}} R_{t} \right)\right. \\
        & p(\nu_{t} | x_{1:(t-1)} , y_{1:(t-1)}) = IG \left(\nu_{t} \left|  \frac{r_{t}}{2} , \frac{r_{t} s_{t-1}}{2} \right) \right. \\
    \end{split}
\end{equation*}
where $IG(a,b)$ denotes inverse-gamma distribution. Sufficient statistics in the above equations are calculated recursively as follows;
\begin{equation*}
\begin{split}
    \text{posterior to prior}&\\
    a_t &= m_{t-1} \ , \  R_t = \frac{1}{\delta} C_{t-1} \ , \  r_t = \beta n_{t-1} \\
    \text{prior to posterior}&\\
    m_t &= a_t + A_t \ , \  C_t = (R_t - A_t A_t^\prime q_t)z_t \ , \ n_t = r_t + 1 \ , \  s_t = s_{t-1}z_t \\
    \text{where} &\quad q_t = s_{t-1} + F_t ^\prime R_t F_t \ , \  z_t = (r_t + (y_t - F_t ^\prime a_t)^2/q_t)/(r_t + 1) \ , \ A_t = R_t F_t / q_t
\end{split}
\end{equation*}
The above procedure is called Forward Filtering (e.g. \citealt{prado2021time}). To implement this model, it is needed to set the initial distribution of the latent variables;
\begin{equation*}
    p(\theta_1 , \nu_1)  =  N \left(\theta_{1} \left| a_{1} , \frac{\nu_{1}}{s_{0}} R_{1} \right)\right. IG \left(\nu_{1} \left|  \frac{r_{1}}{2} , \frac{r_{1} s_{0}}{2} \right) \right. ,
\end{equation*}
and it can be seen that we have to set $(m_0 , C_0 , n_0 , s_0)$. Therefore, the all hyperparameters in this DLM are $(m_0 , C_0 , n_0 , s_0 , \beta , \delta)$.
The predictive distribution is Stundent's t-distribution with the degree of freedom $r_t$;
\begin{equation*}
    p(y_t | x_{1:t},y_{1:(t-1)}) = t(y_t | r_t , f_t , q_t)
\end{equation*}
where $f_t = F_t ^\prime a_t$ is the location parameter and $\sqrt{q_t}$ is the scale parameter. 
The density of this distribution is given by 
\begin{equation*}
    p(y_t | x_{1:t} , y_{1:(t-1)}) = (r_t s_{t-1})^{-\frac{1}{2} } \  B\left( \frac{1}{2}  ,  \frac{r_t}{2}  \right) ^{-1} \left( 1 + \frac{1}{s_{t-1}} F_t^\prime R_t F_t  \right)^{-\frac{1}{2}} \left( 1 + \frac{1}{r_t} \frac{(y_t - a_t ^\prime F_t)^2}{s_{t-1} + F_t ^\prime R_t F_t} \right) ^{-\frac{1+r_t}{2}},
\end{equation*} 
where $B(a,b)$ is the beta function. Conditional on forecasters $x_{1:T}$, all online posterior, online prior, and predictive distributions are computed analytically by Forward Filtering.

The offline posterior distribution $p(\theta_{1:t} , \nu_{1:t} | y_{1:t} , x_{1:t})$ can be approximated by particles. Forward Filtering and Backward Sampling (FFBS) algorithm (e.g., see \citealt{prado2021time}) generates independent particles directly from this distribution. FFBS algorithm at time $T$ is summarized as follows;
\begin{itemize}
    \item[1.]\textbf{Forward Filtering:} Calculate sufficient statistics $(m_t , C_t , n_t , s_t)$ for $t = 1:T$ in sequence.
    \item[2.]\textbf{Backward sampling:} Sample $\Phi_T^{1:N}$ from the online posterior distribution;
    \begin{equation*}
        \begin{split}
            \nu_T^n &\sim IG\left(\frac{n_T}{2} , \frac{n_T s_T}{2}\right) \\
            \theta_T^n &\sim N\left(m_T , \frac{\nu_T^n}{s_T}C_T\right) \ \ \text{for } n = 1:N,
        \end{split}
    \end{equation*}
    and after that, simulate $\Phi_t^{1:N}$ for $t = T-1 : 1$ as follows;
    \begin{equation*}
        \begin{split}
            ( \nu_t^n )^{-1} &= \beta (\nu_{t+1}^n) ^{-1} + \gamma_t^n \ , \ \gamma_t^n \sim Ga\left( (1-\beta)\frac{n_t}{2} , \frac{n_t s_t}{2} \right) \\
            \theta_t^n &\sim N\left( m_t + \delta (\theta_{t+1}^n - m_t) , (1-\delta)\frac{\nu_t^n}{s_t}C_t \right)\ \  \text{for } n = 1:N,
        \end{split}
    \end{equation*}
\end{itemize}
where $Ga(a,b)$ denotes gamma distribution. Particles $\Phi_{1:T}^{1:N}$ generated by the FFBS algorithm approximate the offline posterior distribution.

Since agent predictive distributions are all Student's t-distributions, they can be expressed as the scale mixture of normal distributions;
\begin{equation*}
\begin{split}
        h_{kt}(x_{kt}) &= t(x_{kt} | e_{kt} , \mu_{kt} , H_{kt})\\
        &= \int N(x_{kt} | \mu_{kt} , \sigma_{kt}^2H_{kt}) IG\left(\sigma_{kt}^2 \left| \frac{e_{kt}}{2} , \frac{e_{kt}}{2}\right)\right. d\sigma_{kt}^2
\end{split}
\end{equation*}
where $e_{kt}$ is the degree of freedom, $\mu_{kt}$ is the location parameter, and $\sqrt{H_{kt}}$ is the scale parameter. In this setting, we can implement Gibbs Sampling at time $T$ as follows;
\begin{equation*}
    \begin{split}
        \Phi_{1:T}  &\sim  p(\Phi_{1:T} | x_{1:T} , y_{1:T})\ \  \text{by FFBS} \\
        x_{t}  &\sim N(\mu_t + b_t c_t , H_t - b_t b_t^\prime g_t) \ \ \text{for} \ t = 1:T \\
        \sigma_{kt}^2  &\sim IG\left( \frac{e_{kt}+1}{2} , \frac{e_{kt}+d_{kt}}{2} \right)  \ \ \text{for} \ k = 1:K , t = 1:T ,
    \end{split}
\end{equation*}
where $H_t = \text{diag}(\sigma_{1t}^2H_{1t} , \cdots , \sigma_{Kt}^2H_{Kt}) , \mu_t = (\mu_{1t} , \cdots , \mu_{Kt})^\prime , c_t = y_t - \theta_{t0} - \mu_t^\prime \theta_{t,1:K} , g_t = \nu_t + \theta_{t,1:K}^\prime H_t \theta_{t,1:K} , b_t = H_t \theta_{t,1:K}/g_t$ and $d_{kt} = (x_{kt} - \mu_{kt})^2 / H_{kt}$.

\section{Additional Results for Section~4} \label{app:BPS}

\subsection{Data description}
\begin{figure}[htbp]
\centering
\includegraphics[width=15cm]{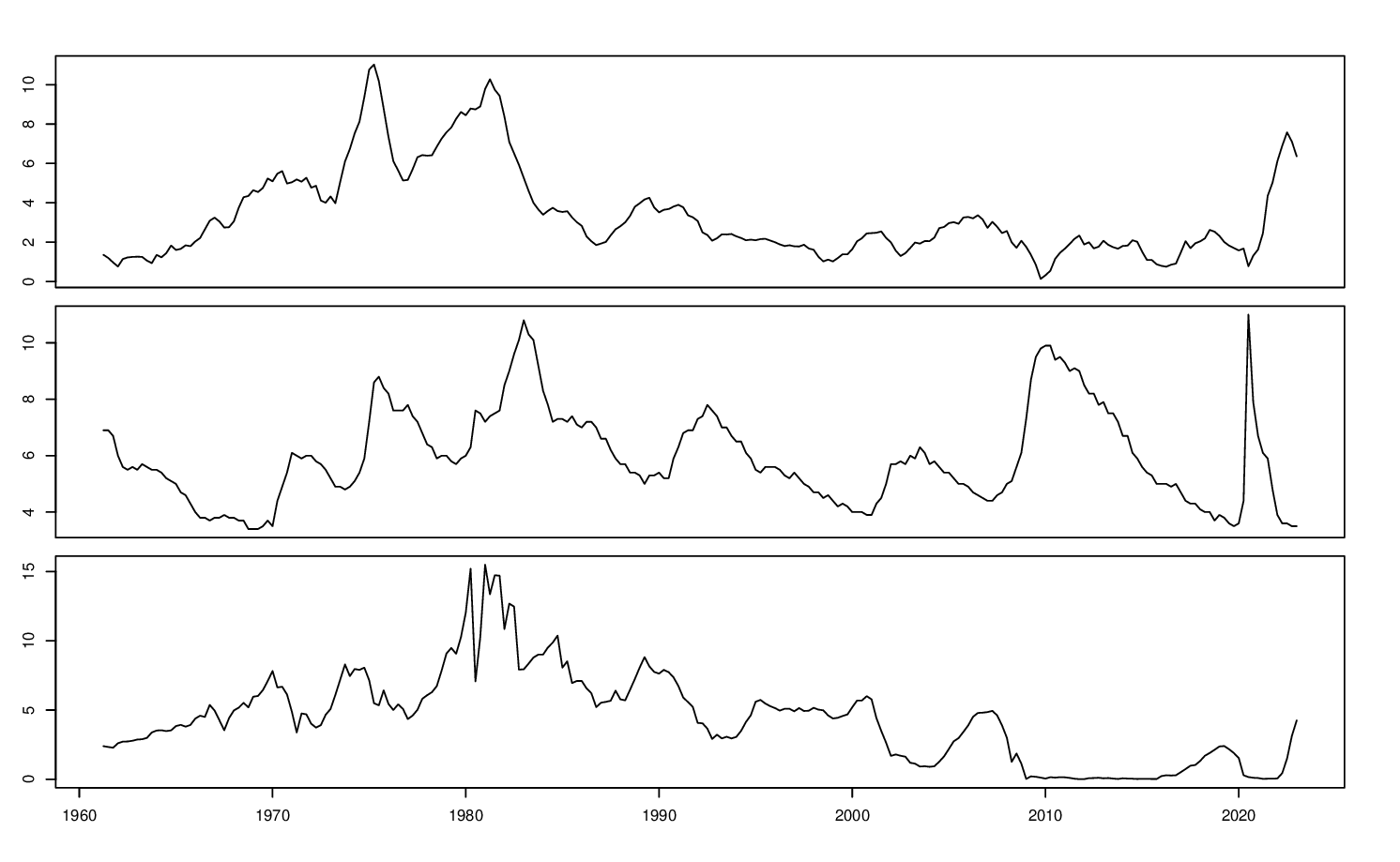}
\caption{US inflation rate (top), unemployment rate (middle) and interest rate (bottom).} \label{fig:data}
\end{figure}

The data on the US inflation, unemployment, and interest rates is retrieved at the webpage of the Federal Reserve Bank of St.~Louis (\url{https://research.stlouisfed.org/series/}). The unemployment and interest rates are recorded monthly, so we create the quarterly data by using the values of the last month in each quarter. Figure~\ref{fig:data} shows the time series of the three variables.

\subsection{Additional posterior plots}
\subsubsection*{Comparison with the MCMC method}
\begin{figure}[htbp]
\centering
\includegraphics[width=15cm]{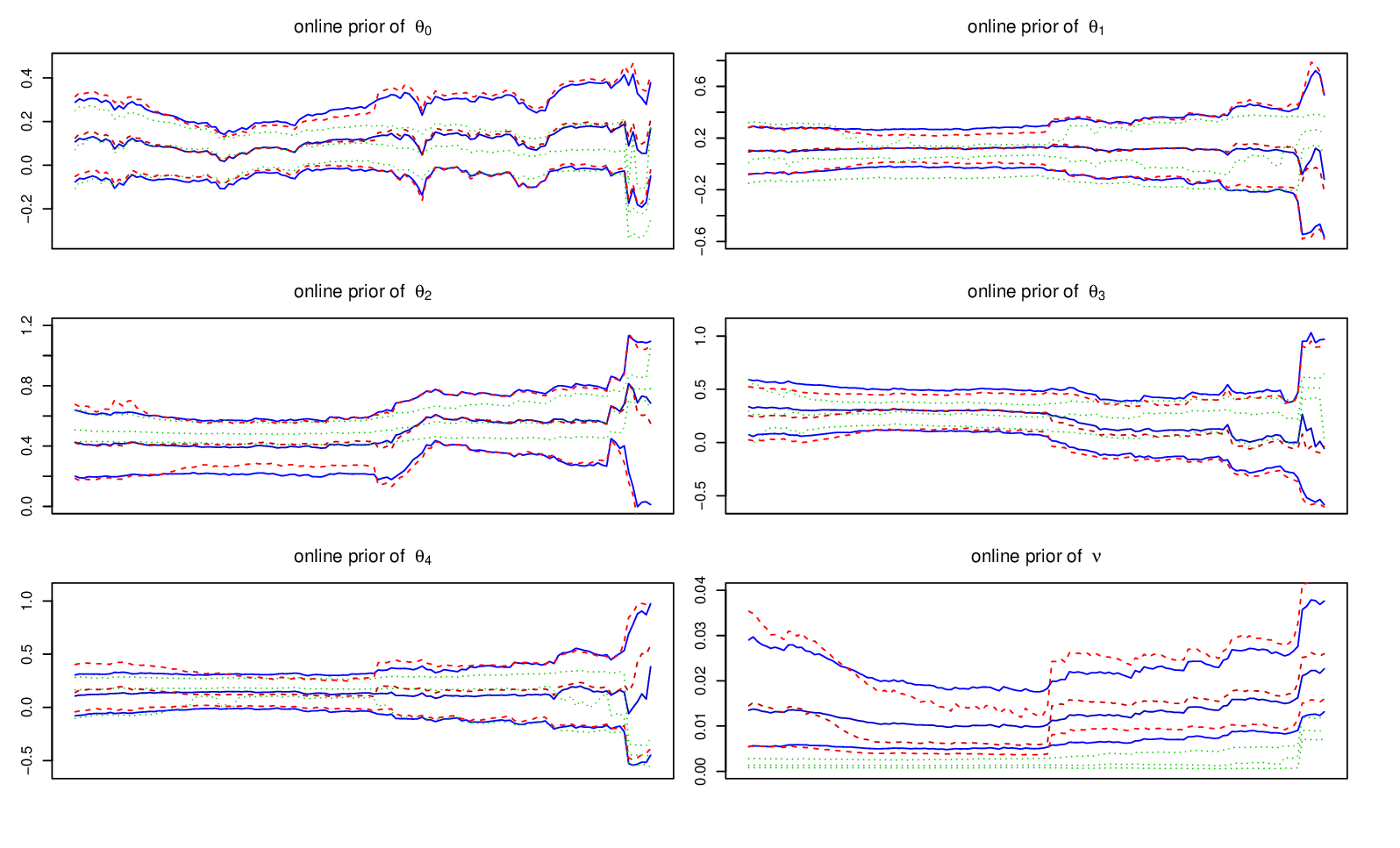}
\caption{The $5\%, 50\%,$ and $95\%$ quantiles of posterior distributions $p(\Phi_t|y_{66:t-1},\mathcal{H}_{66:t-1})$ computed by MCMC (solid), SMC without interventions (dotted) and SMC with interventions (dashed).} \label{fig:smc_marginal}
\end{figure}

In the main text, we illustrated the approximation accuracy of the proposed SMC method by confirming that the {\it marginal} posteriors computed by the SMC and MCMC methods become similar. More marginal plots are shown in Figure~\ref{fig:smc_marginal}, showing the accuracy of the proposed SMC method with interventions. The posteriors approximated without MCMC interventions are biased, especially in estimating variance $\nu_t$, and tend to underestimate the posterior uncertainty. 

\begin{figure}[htbp]
\centering
\includegraphics[width=15cm]{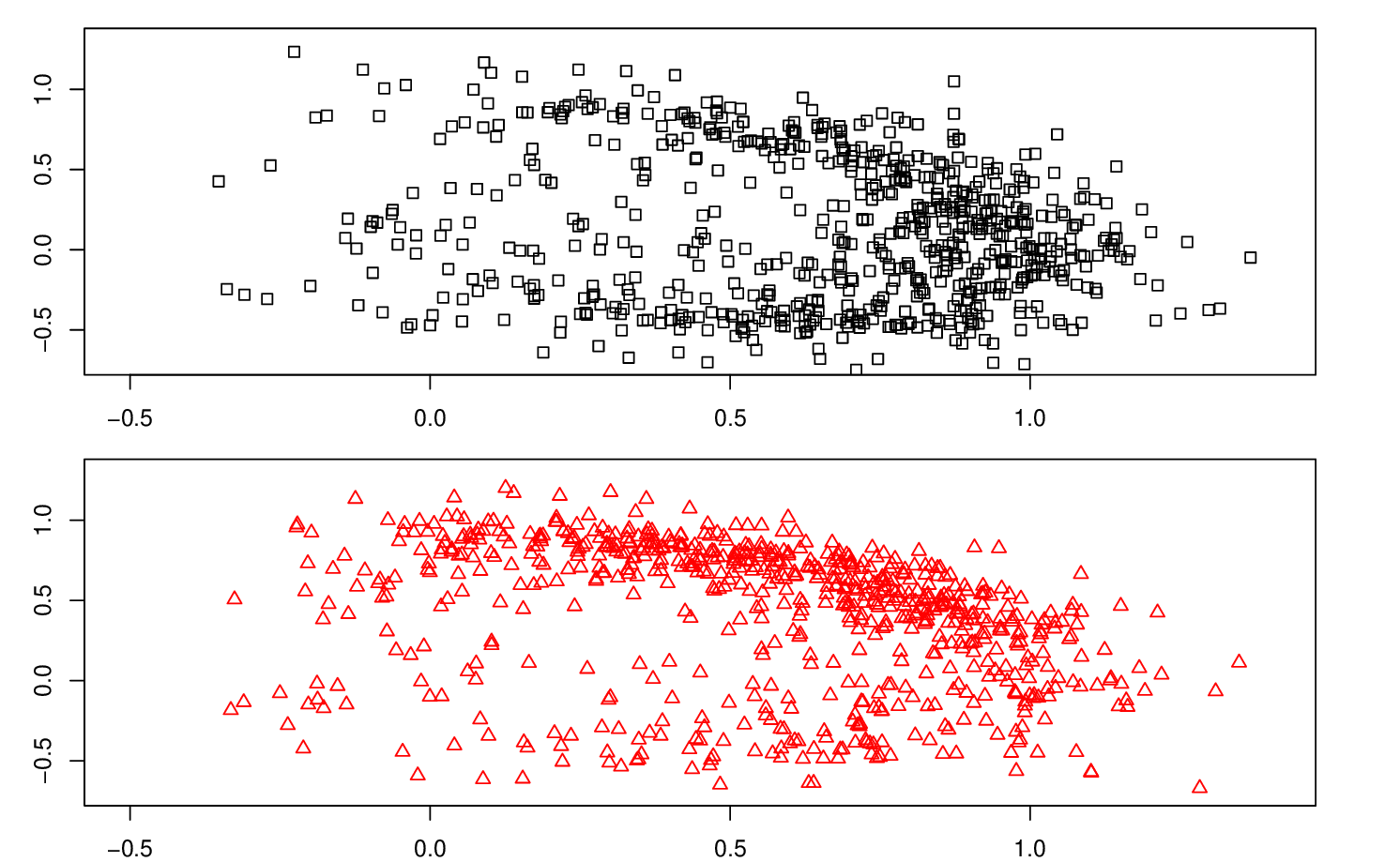}
\caption{Scatter plots of the joint posterior distribution $p(\theta_{2t} , \theta_{4t}|y_{66:t-1},\mathcal{H}_{66:t-1})$ for $t = 247$ computed by MCMC (top) and SMC with interventions (bottom). The horizontal axis is $\theta_{2t}$ and the vertical axis is $\theta_{4t}$.} \label{fig:twoway}
\end{figure}

Here, we show in Figure~\ref{fig:twoway}, the two-dimensional scatter plots of generated $(\theta _{2t},\theta_{4t})$ at $t=247$. These {\it joint} posterior distributions are skewed in a similar way, which supports our claim on the accuracy of the SMC method.

\subsubsection*{Posteriors of $\theta_t$}
\begin{figure}[htbp]
\centering
\includegraphics[width=15cm]{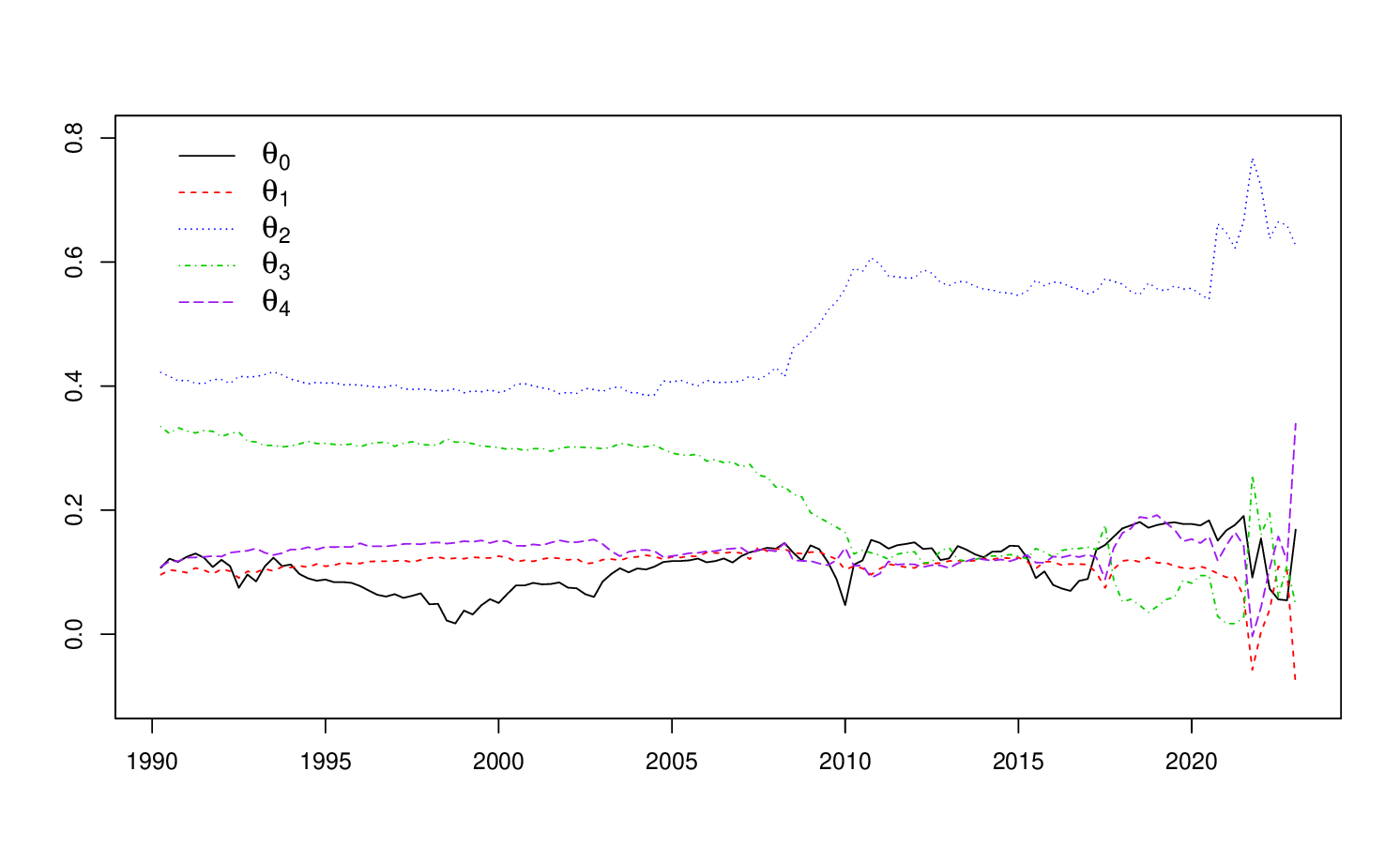}
\caption{Means of $p(\theta_t|y_{66:t-1},\mathcal{H}_{66:t-1})$ for $t = 117:248$ computed by MCMC $(N = 70000)$. The posterior mean of $\theta_{3t}$ decreases gradually in 2005-2010, while $\theta_{2t}$ becomes more dominant in the same period.} \label{fig:coef}
\end{figure}

Figure~\ref{fig:coef} shows the posterior means of calibration coefficients $\theta_t$. Note that this is not a weight vector, so does not sum to unity and can be negative. The highest coefficient is $\theta_{2t}$, the coefficient of $\mathcal{M}_2$ that is most flexible with 9 covariates. In 2020-2022, the coefficient of $\theta_{4t}$ increases in response to the sudden burst of inflation rates. 
Also, in the same period, the intercept, $\theta_{0t}$, frequently changes to adjust the increased predictive bias. 

\subsection{Rough estimate of raw computational time}

We can estimate the computational time of the SMC/MCMC methods by counting the number of parameters to be sampled. Remember that we start the BPS at $t=66$ until $t=T=248$, and use $M$ particles in the SMC method and the chain of length $N$ in the MCMC method. In implementing the Gibbs Sampler, we generate three latent variables $\{x_{t} , \Phi_t , \sigma_t^2 \}$ at each time $t = 66:T$ where $\sigma_t^2 = (\sigma_{t,1}^2 , \cdots , \sigma_{t,K}^2)$. Thus, the total number of generated particles in the Gibbs sampler is $3(T-65)N$. In the SMC method, at each time $t$, we need only $M$ particles to obtain the posterior at time $t$. Then, we estimate the computational time in MCMC by the ratio of the numbers of particles; the MCMC method takes $3(T-65)N/M$ times longer than the SMC method. 

To see this estimate in the real application, consider the actual raw computational times at $T=248$ with particle size $N=M=10000$, which is also reported in Figure~\ref{fig:comptime}. The SMC method takes $0.53$ seconds to complete, while the MCMC method takes $365.5$ seconds, or $\log(363.5) \approx 5.89$ in the scale of log-seconds. The computational time to implement the MCMC method is estimated as $\log( (248-65) \times 3) + \log(0.53) \approx 5.67$, which is close to the actual computational time. This estimate could be useful to infer about the computational time for the MCMC method before its implementation and to decide the number of iterations. However, note that both methods can be speed-up by utilizing parallel computation. 

\subsection{Thresholds of ESS for MCMC intervention}
The intervention by the MCMC method occurs when $\mathrm{ESS}_t < C$. In the main text, we set $C=500$. Here, for different values of threshold $C$, we show the number of interventions, raw computational time and LPDRs in Table~\ref{tab:ESS}. As easily expected, the higher the threshold $C$ is, the more interventions are made, and the more computational time is needed. It is interesting to see that both $C=100$ and $C=250$ have two interventions, but $C=100$ takes longer to compute the posteriors. This is because the interventions occur at later time points (large $t$) when using $C=100$, meaning that the MCMC method is applied to the longer time series and needs to sample many latent variables. The patterns of LPDRs are difficult to interpret as the function of $C$, for the timing of interventions could abruptly change as $C$ increases/decreases. Hence, choosing an optimal $C$ is difficult and requires manual, trial-and-error learning.

\begin{table}[htbp]
    \centering 
    \begin{tabular}{|c|c|c|c|c|c|c|c|} \hline
     $C = $ &  
     \begin{tabular}{c}
     the number of \\
     interventions
     \end{tabular}
     &
    \begin{tabular}{c}
    computational \\
    time (sec)
     \end{tabular} & LPDR $(t = 200)$  & LPDR $(t = 248)$  \\ \hline
    $100$    &  2 & 831.53     & -1.045 & -2.367  \\
    $250$  & 2  & 535.21  & -0.622  & 1.176 \\
    $500$    & 5 & 1266.79  & 0.875 & 1.279   \\
    $1000$    & 9  & 2456.86 & 0.190 & -0.717   \\
    $2000$    & 25  & 6316.75  & -0.019 & 0.042   \\ \hline 
  \end{tabular}
  \caption{Effective sample sizes (ESSs), raw computational time (in seconds) and log predictive density ratios (LPDRs) against the MCMC method $(N = 70000)$ at two points for different values of intervention threshold $C$. The raw computational time with $C = 100$ is longer than that with $C = 250$ because, with $C=100$, the intervention occurs at a later time, which makes computational more costly.}
    \label{tab:ESS}
\end{table}

\section{LDF for BPS ; Real Data analysis setting}\label{app:LDF}

\subsection{Discount factors and LDPLs}
In the real data analysis, we use $J=35$ different values for the DBPS discount factors, $(\beta, \delta)$. Those 35 values are listed as set $S$ below: 
\begin{equation*}
\begin{split}
    S = &\left\{ (r,r) | r = 0.99 - i \cdot 0.01 \  \text{for} \ i = 1:20 \right\} \\
    &\cup \left\{ (r , r - 0.02) | r = 0.99 - i \cdot 0.02 \  \text{for} \ i = 0:4 \right\}\\
    &\cup \left\{ (r , r - 0.04) | r = 0.99 - i \cdot 0.02 \  \text{for} \ i = 0:4 \right\} \\
    &\cup \left\{ (r , r - 0.06) | r = 0.99 - i \cdot 0.02 \  \text{for} \ i = 0:4 \right\} .
\end{split}
\end{equation*}
For each $(\beta _j, \delta _j) \in S$, we can compute the synthesized predictive density $p(y_s | y_{1:s-1} , \mathcal{H}_{1:s} , \beta , \delta)$ for $s = 66:t$ and the LDPL at time $t$ by 
\begin{equation*}
    \text{LDPL}_{j,t}(\gamma) = \sum_{s = 66}^t \gamma ^{t-s} \log p(y_s | \mathcal{H}_{1:s} , y_{1:(s-1)} , \beta _j, \delta _j) ,
\end{equation*}
where $\gamma$ is the LDF discount factor and set as $\gamma = 0.98$ in the main text. We combine the 35 synthesized predictions based on their LDPLs and obtain our prediction for $y_{t+1}$. 

For each $\text{LDF}_{\ast_1,\ast_2}$, we compute the LDPLs in the period of $s = 66:t$. For the first layer discount factor, $\gamma_1$, we consider the following 15 values:
\begin{equation*}
   \gamma_1 \in \{0.01 , 0.3 , 0.5 , 0.6 , 0.7 , 0.75 , 0.8 , 0.85 , 0.9 , 0.92 , 0.95 , 0.97 , 0.98 , 0.99 , 1\} .
\end{equation*}
The second layer discount factor is set as $\gamma _2 = 0.98$.

\subsection{Results about other DBPSs in the LDF}
\begin{figure}[htbp]
\centering
\includegraphics[width=12cm]{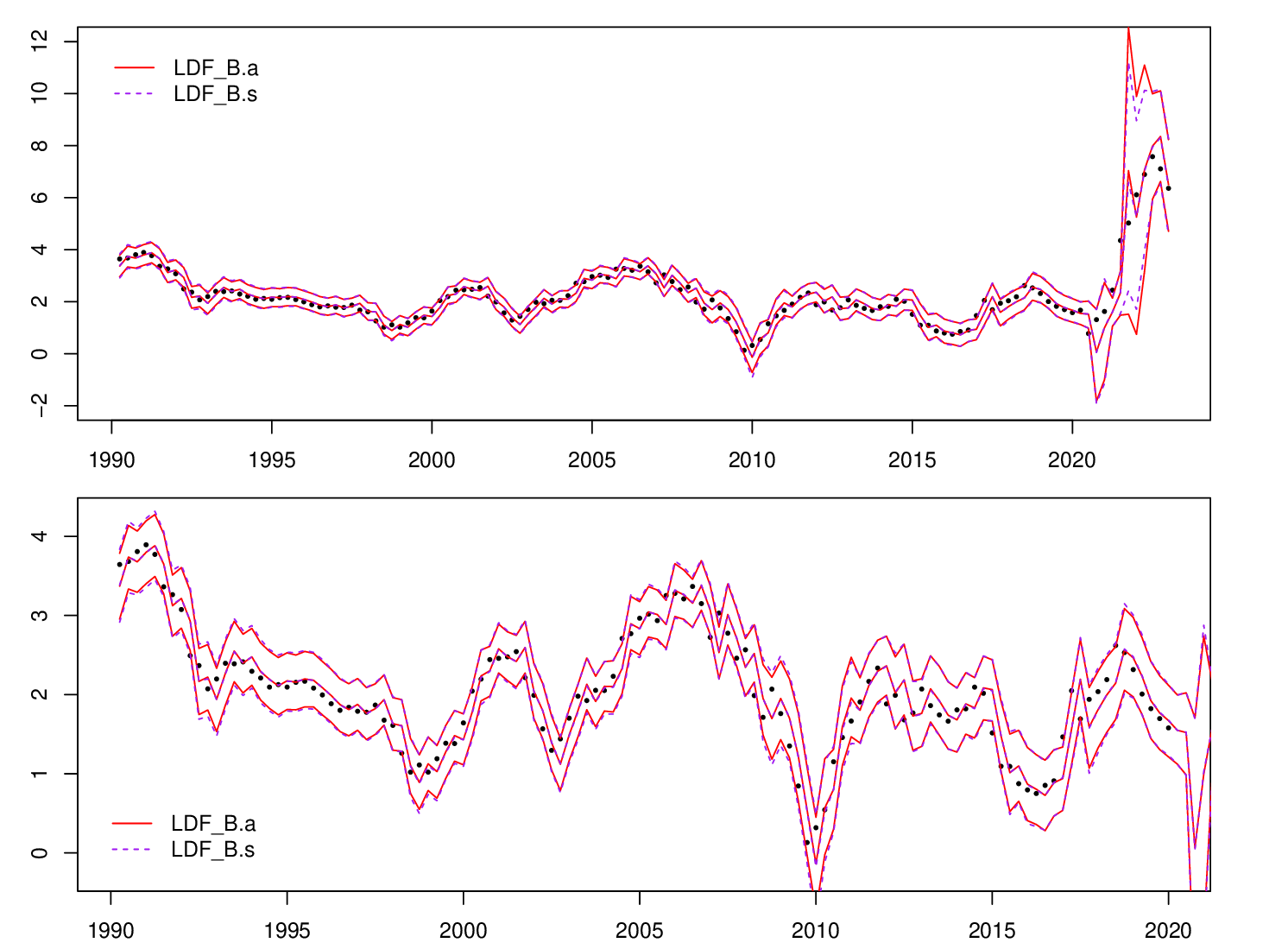}
\caption{
Predictive medians and 90\% predictive intervals of $\text{LDF}_{B,a}$ and $\text{LDF}_{B,s}$ with $\gamma = 0.98$ laid over the data in the period $t = 117:248$ (top) and $t = 117:236$ (bottom). After the sudden structural change at $t = 242$, the $\text{LDF}_{B,a}$ has wider predictive distributions due to the low discount factor it uses. $\text{LDF}_{B,s}$ has slightly wider predictive distributions before that change.} \label{fig:pred_LDF_BPS}
\end{figure}

The predictive distributions of $\text{LDF}_{B,a}$ and $\text{LDF}_{B,s}$ are shown in Figure~\ref{fig:pred_LDF_BPS}. Both DBPSs provide similar predictive distributions. A slight difference can be seen, for example, in the narrower predictive intervals of the $\text{LDF}_{B,a}$ before 2020. This difference can be explained by the use of the argmax function; the predictive distribution of $\text{LDF}_{B,a}$ uses a single discount factor.

\begin{figure}[htbp]
\centering
\includegraphics[width=15cm]{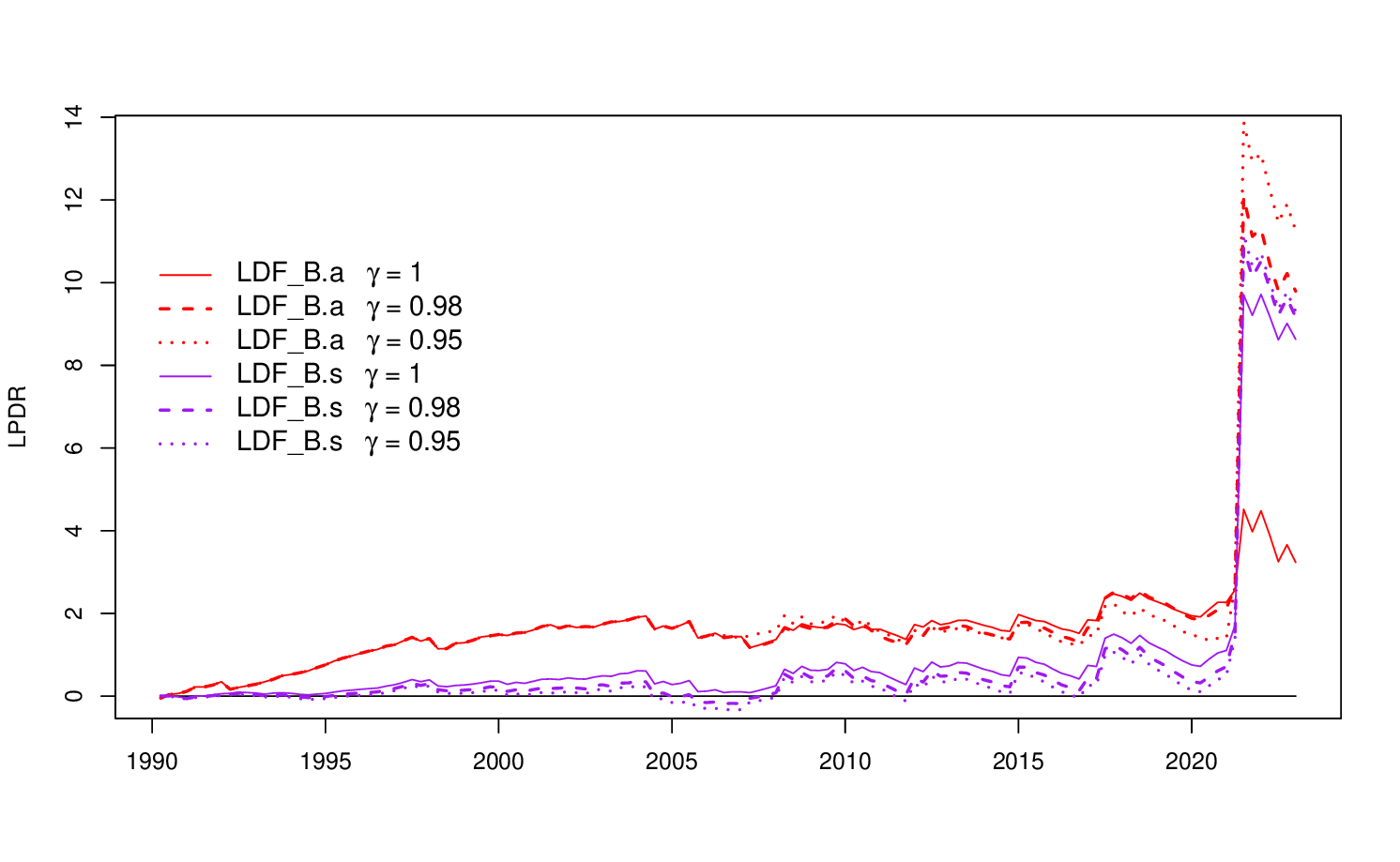}
\caption{Log predictive density rations (LPDRs) of $\text{LDF}_{B,a}$ and $\text{LDF}_{B,s}$ with LDF discount factor $\gamma = 0.95 , 0.98 , 1$ again the benchmark DBPS with the fixed discount factors.} \label{fig:LPDR_BPS}
\end{figure}

We consider three values for the LDF discount factors: $\gamma \in \{ 0.95, 0.98 , 1\}$. Figure~\ref{fig:LPDR_BPS} shows the LPDRs of $\text{LDF}_{B,a}$ and $\text{LDF}_{B,s}$ with three discount factors. If $\gamma$ is high, then the prediction is less adaptive to the sudden change after 2020. If $\gamma$ is low, then the predictions become myopic and less performative before 2020. We conclude that higher values of $\gamma$, but not unity, are preferred.

\begin{figure}[htbp]
\centering
\includegraphics[width=12cm]{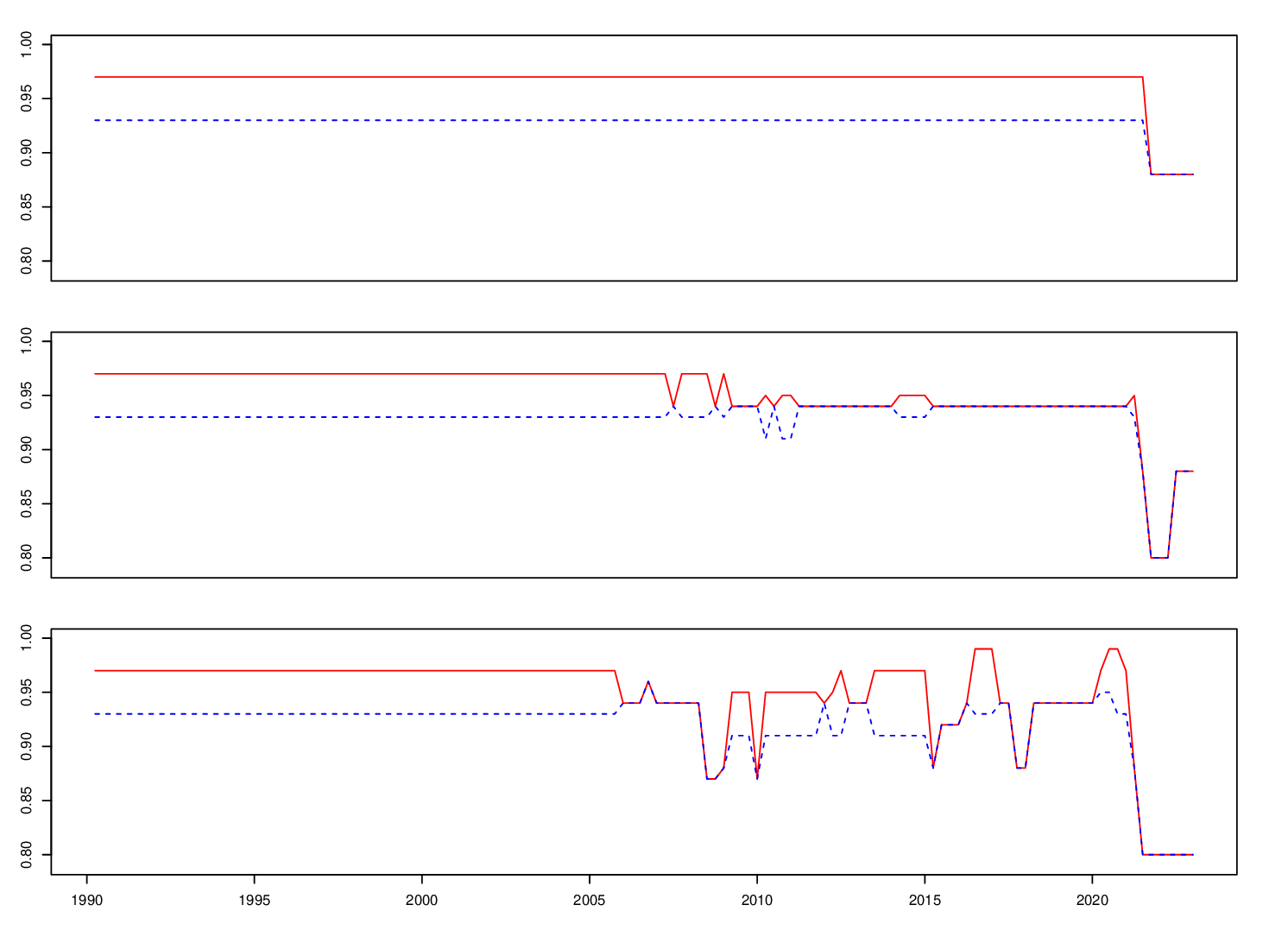}
\caption{The BPS discount factors, $\beta$ (solid) and $\delta$ (dashed), selected in the $\text{LDF}_{B,a}$ by having the highest LDPLs with the LDF discount factor $\gamma = 1$ (top), $\gamma = 0.98$ (middle) and $\gamma = 0.95$ (bottom).} \label{fig:DBPS_df}
\end{figure}

Figure~\ref{fig:DBPS_df} shows the values of BPS discount factors $\beta$ and $\delta$ that have the highest LDPLs in the $\text{LDF}_{B,a}$. In response to the sudden burst, both discount factors to adjust bias in predictive location and uncertainty. Furthermore, in cases of $\gamma = 0.98$, the value of the discount factor increases again after the burst, reducing the predictive uncertainty. 

\subsection{Results about other loss discounting approaches}

\begin{figure}[htbp]
\centering
\includegraphics[width=15cm]{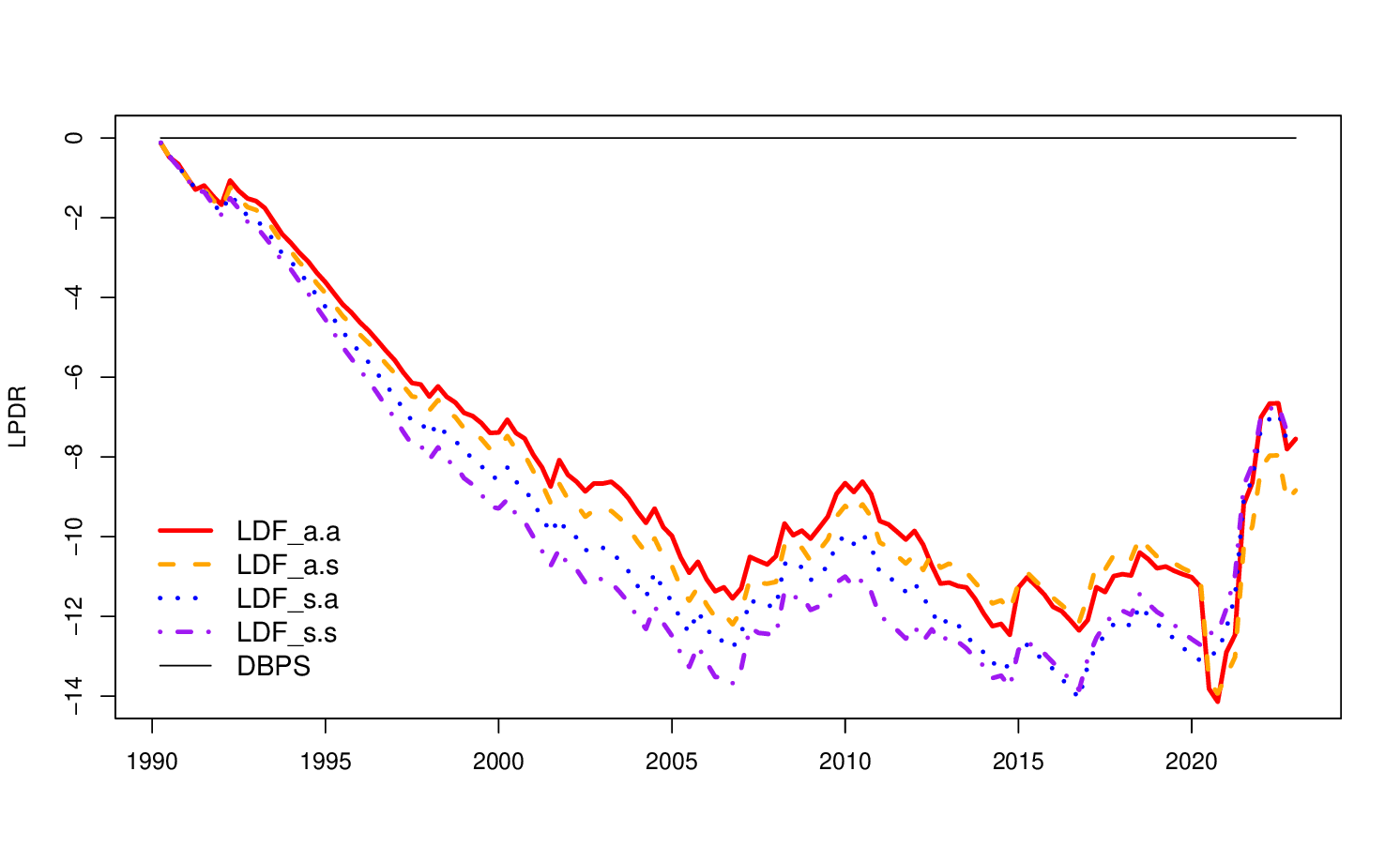}
\caption{LPDRs of the four LDFs with $\gamma _2= 0.98$.}\label{fig:LPDR_LDF}
\end{figure}

Figure~\ref{fig:LPDR_LDF} shows the LPDRs computed in the LDF for four combinations of the softmax and argmax weights. The $\text{LDF}_{s,a}$ used in the main text performs satisfactorily both before and after 2020. All four models are not as competitive as the benchmark DBPS. 

\begin{figure}[htbp]
\centering
\includegraphics[width=12cm]{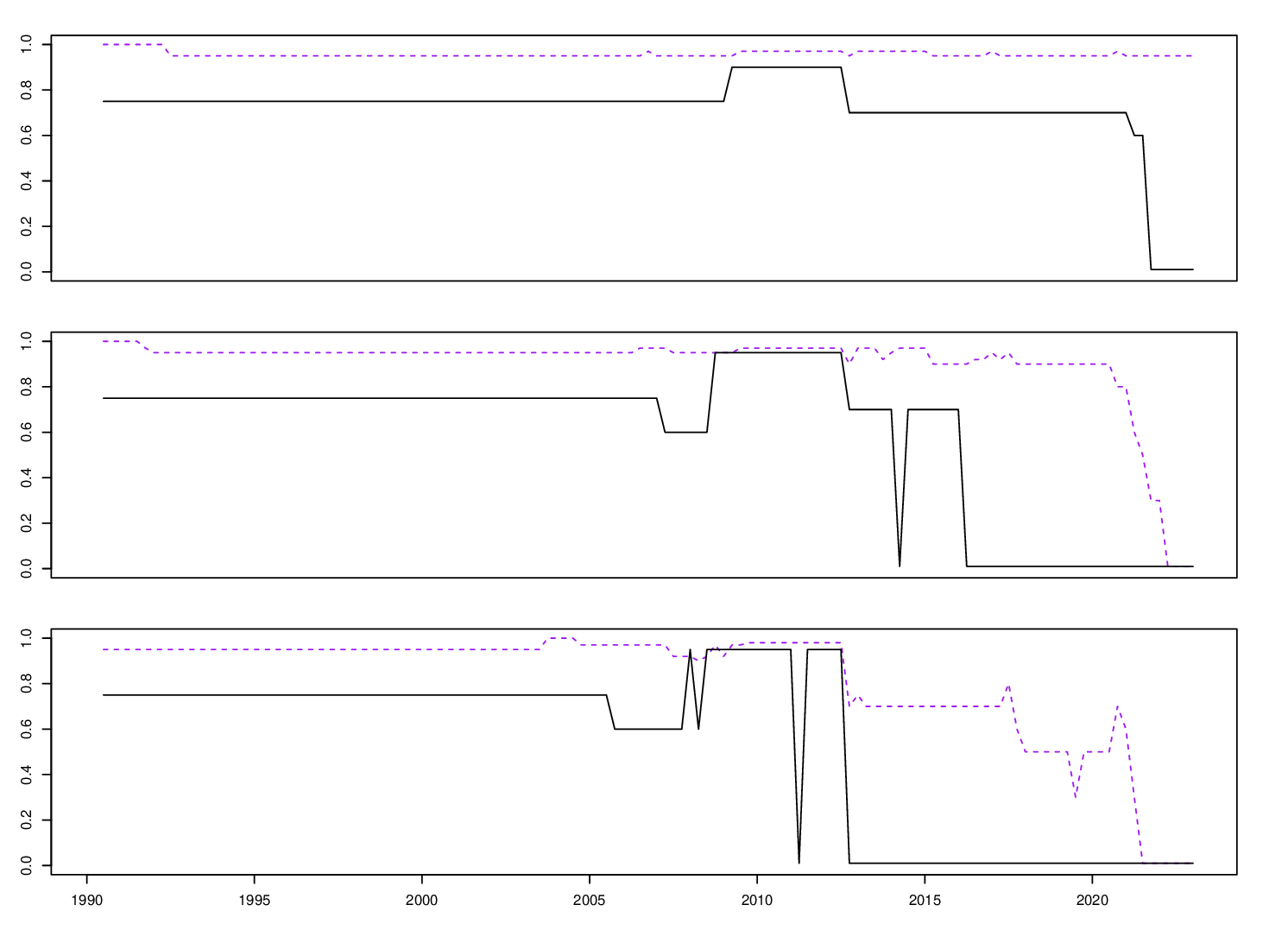}
\caption{
The first layer discount factors, $\gamma_1$, selected in the $\text{LDF}_{a,a}$ (solid) and $\text{LDF}_{s,a}$ (dashed) by having the highest LDPLs with the LDF discount factor $\gamma_2 = 1$ (top), $\gamma_2 = 0.98$ (middle) and $\gamma_2 = 0.95$ (bottom). } \label{fig:LDF_df} 
\end{figure}

The first layer discount factor, $\gamma_1$, that has the highest LDPL is shown in Figure~\ref{fig:LDF_df}. The sudden decrease in the discount factor can also be seen in the LDF. In some cases, however, this decrease occurs earlier than 2020, responding to fluctuations in inflation rates in 2013-2016.

\end{document}